\documentclass{article}

\usepackage{graphicx}
\usepackage{multirow}
\usepackage{amssymb}
\usepackage{amsthm}
\usepackage{amsmath}
\usepackage{fullpage}
\usepackage[longnamesfirst]{natbib}
\usepackage{enumerate}
\usepackage{amsfonts, bm, amsmath, color, natbib, rotating, amsthm, subfigure, lscape, multicol, epstopdf, verbatim, hyperref}
\usepackage{authblk}
\usepackage{xr}

\usepackage{booktabs}

\bibpunct{(}{)}{;}{a}{,}{,}

\usepackage{setspace}
\theoremstyle{plain} \newtheorem{theorem}{Theorem} \newtheorem{proposition}{Proposition} \newtheorem{lemma}{Lemma} 

\theoremstyle{definition}    \newtheorem{algorithm}{Algorithm}
\newcommand{\utwi}[1]{\mbox{\boldmath $ #1$}}
\theoremstyle{remark}   

\begin{document}

\newif\ifblinded

\title{Bayesian Function-on-Scalars Regression for High Dimensional Data}

\ifblinded
\author{}
\else

\author{Daniel R. Kowal and Daniel C. Bourgeois\thanks{Kowal is Assistant Professor, Department of Statistics, Rice University, Houston, TX 77251-1892 (E-mail: \href{mailto:Daniel.Kowal@rice.edu}{Daniel.Kowal@rice.edu}). Bourgeois is a PhD student, Department of Statistics, Rice University, Houston, TX 77251-1892 (E-mail: \href{mailto:Daniel.C.Bourgeois@rice.edu}{Daniel.C.Bourgeois@rice.edu})}
\thanks{This material is based upon work supported by the National Science Foundation Graduate Research Fellowship (Bourgeois) under Grant DGE-1450681
and partially supported by NSF award \#1547433 RTG: Cross-training in Statistics and Computer Science.}}
\fi

\maketitle
\large

\vspace{-10mm}

\begin{abstract}
We develop a fully Bayesian framework for function-on-scalars regression with many predictors. The functional data response is modeled nonparametrically using unknown basis functions, which produces a flexible and data-adaptive functional basis. We incorporate shrinkage priors that effectively remove unimportant scalar covariates from the model and reduce sensitivity to the number of (unknown) basis functions. For variable selection in functional regression, we propose a decision theoretic posterior summarization technique, which identifies a subset of covariates that retains nearly the predictive accuracy of the full model. Our approach is broadly applicable for Bayesian functional regression models, and unlike existing methods provides joint rather than marginal selection of important predictor variables. Computationally scalable posterior inference is achieved using a Gibbs sampler with linear time complexity in the number of predictors. The resulting algorithm is empirically faster than existing frequentist and Bayesian techniques, and provides joint estimation of model parameters, prediction and imputation of functional trajectories, and uncertainty quantification via the posterior distribution. A simulation study demonstrates improvements in estimation accuracy, uncertainty quantification, and variable selection relative to existing alternatives. The methodology is applied to actigraphy data to investigate the association between intraday physical activity and responses to a sleep questionnaire.





\end{abstract}

\clearpage

\section{Introduction}
Modern scientific measuring systems commonly record data over a continuous domain, often at high resolutions, and referred to as \emph{functional data}. Functional data are typically high dimensional and highly correlated, and may be measured concurrently with other variables of interest. We consider the problem of \emph{function-on-scalars regression} (FOSR), in which a functional data response is modeled using (possibly many) \emph{scalar} predictors. Applications of FOSR are broad and impactful: examples include blood pressure profiles during pregnancy \citep{montagna2012bayesian}, human motor control following a stroke \citep{chen2016variable}, age-specific fertility rates \citep{kowal2018dynamic},
microarray time course gene expression data \citep{wang2007group}, and longitudinal genome-wide association studies  \citep{barber2017function,fan2017high}, among others \citep{silverman2005functional, morris2015functional}.

FOSR shares the same fundamental goals as multiple regression---estimation and inference for the regression coefficients, selection of important predictor variables, and prediction of new responses---with additional modeling challenges. Within-curve dependence of functional data requires careful modeling of the covariance function, which may be complex, with implications for computational scalability. In addition, the regression coefficients in FOSR are \emph{functions}, which complicates estimation, inference, and selection. Naturally, these challenges are compounded in high-dimensional settings. Lastly, prediction in FOSR requires prediction of functional trajectories, which may be partially observed.

To address these challenges, we develop a fully Bayesian framework for FOSR, where the number of predictors $p$ may be greater than the number of observed curves $n$. Within-curve dependence is modeled nonparametrically using \emph{unknown} basis functions, which produces a data-adaptive functional basis with uncertainty quantification via the posterior distribution. We introduce shrinkage priors to mitigate the impact of unimportant predictor variables and reduce sensitivity to the number of basis functions. For computationally scalable posterior inference, we develop a Gibbs sampling algorithm with linear time complexity in either $n$ or $p$.
%
The methodology is applicable for densely- or sparsely-observed functional data, with automatic prediction and imputation at unobserved points within the Gibbs sampler.

Recent developments in FOSR have focused on variable selection, which requires selecting or thresholding entire regression coefficient \emph{functions}. Accordingly, various group penalties have been proposed for variable selection in FOSR, including a group smoothly clipped absolute deviation (\citealp{wang2007group}) and a group-minimix concave penalty (\citealp{chen2016variable}) for a moderate number of predictors, and a group lasso \citep{barber2017function} and an adaptive group lasso \citep{fan2017high} for high-dimensional predictors. Although these frequentist methods provide both estimation and selection,  finite-sample inference is unavailable and important tuning parameters must be selected, which is often computationally intensive. By comparison, Bayesian approaches provide exact finite-sample inference (up to MCMC error) and do not require selection of tuning parameters. Bayesian FOSR models may identify important predictors marginally using \emph{global Bayesian $p$-values} (GBPVs; \citealp{meyer2015bayesian}). However, there is currently no coherent Bayesian framework for simultaneous estimation, inference, and variable selection in FOSR, particularly for moderate to large $p$.

We propose a decision theoretic approach for Bayesian variable selection in FOSR. Using a loss function that balances sparsity with predictive accuracy for functional responses, we obtain sparse posterior summaries by optimizing the expected loss under the posterior predictive distribution. As a result, we identify a subset of predictors that maintains nearly the predictive accuracy of the full model. Notably, our approach selects variables \emph{jointly} rather than \emph{marginally}, is applicable under a variety of priors, and can be tractably solved using existing software for penalized regression. The methodology may be viewed as an extension of \cite{hahn2015decoupling} to the functional data setting.

We apply our methods to study the association between intraday physical activity and responses to a sleep questionnaire among elderly adults. It is \emph{a priori} unclear which questionnaire items, if any, are associated with physical activity, as measured by actigraphy data, and whether or not such associations vary throughout the day. The proposed methodology provides the framework to (i) model and impute intraday physical activity trajectories for each individual, (ii) estimate regression coefficient functions and accompanying posterior credible bands,
and (iii) select a small subset of the questionnaire items and demographic variables ($p=74$) that maintain nearly the predictive accuracy of the full model.

The remainder of the paper is organized as follows:
the model is in Section \ref{fosr};
the MCMC algorithm is in Section \ref{algorithm};
the variable selection method is in Section \ref{dss};
a simulation analysis is in Section \ref{simulations};
the application is in Section \ref{app};  we conclude in Section \ref{conclusions}. MCMC diagnostics, proofs of theoretical results, and details on the application are provided in the Appendix. An \texttt{R} package is available on GitHub and example code is available as a supplementary file.

\section{A Bayesian Function-on-Scalars Regression Model}\label{fosr}
\subsection{Model Specification and Assumptions}
Let $Y_1(\tau),\ldots, Y_n(\tau)$ be a functional data response with $\tau \in \mathcal{T}$, where $\mathcal{T} \subset \mathbb{R}^D$ is a compact index set and $D \in \mathbb{Z}^+$. Suppose we have $p$ scalar predictors $\{x_{i,j}\}_{j=1}^p$ for $i=1,\ldots,n$, possibly with $p > n$. We are interested in modeling the association between the scalar predictors $x_{i,j}$ and the functional response $Y_i$. We propose the following  Bayesian FOSR model:
\begin{align}
\label{fda}
Y_i(\tau) &= \sum_{k=1}^K f_k(\tau) \beta_{k,i} + \epsilon_i(\tau),\quad \epsilon_i(\tau) \stackrel{indep}{\sim}N(0, \sigma_\epsilon^2),\quad \tau \in \mathcal{T}  \\
\label{reg}
\beta_{k,i} &= \mu_{k} + \sum_{j=1}^p x_{i,j}  \alpha_{j,k} + \gamma_{k,i}   \\
\label{priors}
\quad \mu_{k}  \stackrel{indep}{\sim} N(0, \sigma_{\mu_k}^2), &\quad \alpha_{j,k}  \stackrel{indep}{\sim} N(0, \sigma_{\alpha_{j,k}}^2), \quad \gamma_{k,i}  \stackrel{indep}{\sim}N(0, \sigma_{\gamma_{k,i}}^2)
\end{align}
Model \eqref{fda} expands the functional data $Y_i(\cdot)$ in the basis $\{f_k(\cdot)\}$ with a regression model for the corresponding basis coefficients $\{\beta_{k,i}\}$. The basis functions $\{f_k(\tau)\}$, which may be known (e.g., splines or wavelets) or unknown (see Section \ref{loadings}), capture within-curve dependence of the functional data $Y_i$, while the coefficients $\{\beta_{k,i}\}$ model between-curve dependence induced by the predictor variables $x_{i,j}$. The intercepts $\{\mu_k\}$, the regression coefficients $\{\alpha_{j,k}\}$, and the subject-specific errors $\{\gamma_{k,i}\}$ are given conditionally Gaussian priors (see Section \ref{shrinkage}), and we assume a Jeffreys' prior for the observation error variance, $\left[\sigma_\epsilon^2 \right]\propto 1/\sigma_\epsilon^2$.

Model \eqref{fda}-\eqref{priors} may be expressed as a more conventional FOSR model. Let $\mathcal{GP}(c, C)$ denote a Gaussian process with mean function $c$ and covariance function $C$.
\begin{proposition} \label{model_equiv}
Model \eqref{fda}-\eqref{priors} implies the functional regression model
\begin{equation}
\label{fosr_fda}
Y_i(\tau) =  \tilde \mu(\tau) + \sum_{j=1}^p x_{i,j}  {\tilde \alpha}_j(\tau) + \tilde \gamma_i(\tau) + \epsilon_i(\tau), \quad \epsilon_i(\tau) \stackrel{indep}{\sim}N(0, \sigma_{\epsilon}^2), \quad \tau \in \mathcal{T}
\end{equation}
with expansions $\tilde \mu(\tau) \equiv \sum_{k=1}^K  f_k(\tau) \mu_k \sim \mathcal{GP}(0, C_\mu)$ for $C_\mu(\tau, u) = \sum_{k=1}^K f_k(\tau) f_k(u) \sigma_{\mu_k}^2$, 
$ {\tilde \alpha}_{j}(\tau)  \equiv \sum_{k=1}^K f_k(\tau)  \alpha_{j,k} \stackrel{indep}{\sim} \mathcal{GP}(0, C_{\alpha_j})$ for  $C_{\alpha_j}(\tau, u) = \sum_{k=1}^K f_k(\tau) f_k(u) \sigma_{\alpha_{j,k}}^2 $, and
$ \tilde \gamma_i(\tau) \equiv \sum_{k=1}^K f_k(\tau) \gamma_{k,i} \stackrel{indep}{\sim}\mathcal{GP}(0, C_{\gamma_{i}})$ for $C_{\gamma_{i}}(\tau, u) = \sum_{k=1}^K f_k(\tau) f_k(u) \sigma_{\gamma_{k,i}}^2$.
\end{proposition}
The predictors $x_{i,j}$ are directly associated with the functional data $Y_i(\tau)$ via the regression coefficient functions $\tilde \alpha_{j}(\tau) =\sum_k f_k(\tau)  \alpha_{j,k}$. The subject-specific error term $\tilde \gamma_i(\tau)$ in \eqref{fosr_fda} captures within-curve variability in $Y_i(\tau)$ unexplained by $\{x_{i,j}\}$, which marginally produces a model for within-curve correlations of the FOSR error. Accounting for within-curve error correlations is important for statistically efficient estimation and valid inference in FOSR \citep{reiss2010fast}, especially for high-dimensional predictors \citep{chen2016variable}.

\subsection{Modeling the Basis Functions}\label{loadings}
For broad applicability, the basis functions $\{f_k\}$ in \eqref{fda} should be flexible, efficient for computations, and well-defined for $\mathcal{T}\subset \mathbb{R}^D$ with $D \in \mathbb{Z}^+$. The dimension $K$ is important: each predictor variable $j=1,\ldots,p$ is accompanied by $K$ coefficients, which may be correlated. FOSR methods that use full basis expansions, such as splines or wavelets, have been successfully applied for small $p$ \citep{silverman2005functional}, including extensions for mixed effects models \citep{guo2002functional, morris2006wavelet,zhu2011robust,goldsmith2016assessing}, but are neither parsimonious nor computationally scalable for moderate to large $p$.
One remedy is to pre-compute a lower-dimensional basis, such as a functional principal components (FPC)  basis. However, this approach implicitly conditions on the estimates FPCs and fails to account for the accompanying uncertainty. \cite{goldsmith2013corrected} find that FPC-based methods can substantially underestimate total variability, even for densely-sampled functional data.

We instead model each $f_k$ as an unknown, smooth function. By modeling $\{f_k\}$ as unknown, we simultaneously  (i) produce a data-adaptive functional basis, which minimizes the number of basis functions $K$ needed, and (ii) incorporate uncertainty quantification of $\{f_k\}$ via the posterior distribution. Smoothness encourages information sharing among nearby points, which reduces variability to produce more stable prediction at unobserved points. Model identifiability is enforced by coupling a matrix orthonormality constraint on $\{f_k\}$ with an ordering constraint on the variance components in \eqref{priors} (see Section \ref{algorithm}).

We adopt the model for $\{f_k\}$ in \cite{kowal2018dynamic}, which offers substantial computational improvements over existing alternatives \citep{montagna2012bayesian,goldsmith2015generalized,suarez2017bayesian}. Let $f_k(\tau) = \bm b'(\tau) \bm \psi_k$ for known basis functions $\bm b'(\tau) = (b_1(\tau), \ldots, b_{L_m}(\tau))$ and unknown basis coefficients $\bm\psi_k$. We use low rank thin plate splines (LR-TPS) for $\bm b(\cdot)$, which are are well-defined for $\mathcal{T}\subset \mathbb{R}^D$ with $D \in \mathbb{Z}^+$ and efficient in MCMC samplers \citep{crainiceanu2005bayesian}. Smoothness is encouraged via the prior $\bm \psi_k \sim N(\bm 0, \lambda_{f_k}^{-1} \bm \Omega^{-1})$, where $\bm \Omega$ is a $L_m \times L_m$ known roughness penalty matrix. The smoothing parameter $ \lambda_{f_k} $ appears as a prior precision, so we assign a uniform prior distribution on the corresponding standard deviation,  $\lambda_{f_k}^{-1/2} \stackrel{iid}{\sim} \mbox{Uniform}(0, 10^4)$ \citep{kowal2017bayesian}. Details on the construction of $\bm b(\cdot)$ and $\bm \Omega$ are given in \cite{kowal2018dynamic} and the relevant full conditional distributions are in Section \ref{algorithm}.

\subsection{Shrinkage Priors}\label{shrinkage}
The regression model \eqref{reg} may include unimportant predictors, especially for moderate to large $p$. Without regularization or shrinkage, such irrelevant predictors can reduce estimation accuracy and statistical efficiency. Model \eqref{fda}-\eqref{priors} also requires a choice of $K$. 
While $K$ may be treated as unknown and estimated in the model, this approach typically requires computationally intensive procedures, such as reversible jump MCMC \citep{suarez2017bayesian}. Instead, we impose ordered shrinkage with respect to $k=1,\ldots,K$, so that larger number factors are \emph{a priori} less important. Ordered shrinkage is computationally scalable and empirically reduces sensitivity to the choice of $K$, provided $K$ is chosen sufficiently large.

We include a groupwise horseshoe prior for the regression coefficients $\alpha_{j,k} \stackrel{indep}{\sim} N(0, \sigma_{\alpha_{j,k}}^2)$, which extends \cite{carvalho2010horseshoe}. Shrinkage is applied at both the factor-within-predictor level as well as the predictor-level using a hierarchy of half-Cauchy distributions:
\begin{equation}\label{shrink}
\sigma_{\alpha_{j,k}} \stackrel{ind}{\sim} C^+(0, \lambda_{j}), \quad \lambda_{j} \stackrel{ind}{\sim} C^+(0, \lambda_{0}), \quad \lambda_{0} \stackrel{ind}{\sim} C^+(0, p^{-1/2})
\end{equation}
The scale parameter $\sigma_{\alpha_{j,k}}$ controls the factor $k$ shrinkage for predictor $j$, $\lambda_j$ determines the shrinkage for all regression coefficients $\{\alpha_{j,k}\}_{k=1}^K$ for predictor $j$, and  $\lambda_0$ corresponds to the global level of sparsity for all predictors $j=1,\ldots,p$, and is scaled by $p^{-1/2}$ following \cite{piironen2016hyperprior}. The horseshoe prior and its variants have been successful in a variety of models and applications, including functional regression \citep{kowal2017dynamic,kowal2018dynamic}.

We apply the multiplicative gamma process (MGP) for ordered shrinkage \citep{bhattacharya2011sparse}. MGP priors for the intercepts $\{\mu_k\}$ and the subject-specific errors $\{\gamma_{k,i}\}$ are represented via priors on the respective variance components in \eqref{priors}. The intercept prior precisions are $\sigma_{\mu_k}^{-2} = \prod_{\ell \le k} \delta_{\mu_\ell}$, where $ \delta_{\mu_1} \sim \mbox{Gamma}(a_{\mu_1}, 1)$ and $ \delta_{\mu_\ell} \sim \mbox{Gamma}(a_{\mu_2}, 1)$ for $\ell > 1$, which implies a stochastic ordering for $\sigma_{\mu_k}^2$ when $a_{\mu_1} > 0$ and $a_{\mu_2} \ge 2$ \citep{bhattacharya2011sparse}.
For the subject-specific errors, we let $\sigma_{\gamma_{k,i}}^2 = \sigma_{\gamma_k}^2/\xi_{\gamma_{k,i}}$ with $\sigma_{\gamma_k}^{-2} = \prod_{\ell \le k} \delta_{\gamma_\ell}$, $ \delta_{\gamma_1} \sim \mbox{Gamma}(a_{\gamma_1}, 1)$, $ \delta_{\gamma_\ell} \sim \mbox{Gamma}(a_{\gamma_2}, 1)$ for $\ell > 1$, and $\xi_{\gamma_{k,i}} \stackrel{iid}{\sim} \mbox{Gamma}(\nu_\gamma/2, \nu_\gamma/2)$, as in \cite{bhattacharya2011sparse} and \cite{montagna2012bayesian}. The hyperpriors  $a_{\mu_1}, a_{\mu_2}, a_{\gamma_1},a_{\gamma_2} \stackrel{iid}{\sim}\mbox{Gamma}(2,1)$ allow the data to determine the rate of ordered shrinkage separately for $\{\mu_k\}$ and $\{\gamma_{k,i}\}$. Lastly, the hyperprior $\nu_\gamma \sim \mbox{Uniform}(2, 128)$ induces heavy tails in the marginal distribution of $\gamma_{k,i}$ for additional model robustness. 




\section{MCMC Sampling Algorithm} \label{algorithm}
We construct an MCMC sampling algorithm that consists of  efficient closed form sampling steps (with the exception of the shrinkage prior hyperparameters $a_{\mu_1}, a_{\mu_2}, a_{\gamma_1},a_{\gamma_2}, \nu_\gamma$, which alternatively may be fixed in advanced). The main blocks of the sampling algorithm are (i) the basis functions $\{f_k\}$ in \eqref{fda}, (ii) the regression coefficients $\{\mu_k, \alpha_{j,k}, \gamma_{k,i}\}$ in \eqref{reg}, and (iii) the variance components in \eqref{fda} and \eqref{priors}. An overview of the algorithm is presented here, with details provided in the Appendix.

Suppose we observe the functional data $Y_i$ at observation points $\{\tau_\ell\}_{\ell=1}^m$.  For notational convenience, we assume the observation points are identical for all subjects $i$, but later relax that assumption  (see Section \ref{app}). The likelihood in \eqref{fda} becomes
\begin{equation}\label{fdaVec}
\bm Y_i = \sum_{k=1}^K \bm f_k \beta_{k,i} + \bm \epsilon_i,
\quad \bm \epsilon_i \sim N(\bm 0, \sigma_{\epsilon}^2 \bm I_m)
\end{equation}
where $\bm Y_i = (Y_i(\tau_1),\ldots,  Y_i(\tau_m))'$ and $\bm f_k =  (f_k(\tau_1),\ldots,  f_k(\tau_m))'$. For functional data, often $m$ is large and the  components of $\bm Y_i$ are highly correlated. Therefore, MCMC sampling algorithms must be constructed carefully to ensure both computational and MCMC efficiency.

First, we sample the unknown basis functions $f_k(\tau) = \bm b'(\tau) \bm\psi_k$ by iteratively drawing from the full conditional distribution of $\bm \psi_k$ given $\{\bm\psi_\ell\}_{\ell\ne k}$: $\left[\bm \psi_k | \cdots \right] \sim N\left(\bm Q_{\psi_k}^{-1} \bm \ell_{\psi_k}, \bm Q_{\psi_k}^{-1}\right)$  for $k=1,\ldots, K$, where
$\bm Q_{\psi_k} =  \sigma_\epsilon^{-2}(\bm B'\bm B)\sum_{i=1}^n \beta_{k,i}^2 + \lambda_{f_k} \bm \Omega$ and  $\bm \ell_{\psi_k} =  \sigma_\epsilon^{-2}\bm B' \sum_{i=1}^n \big[\beta_{k,i} \big( \bm Y_i - \sum_{k' \ne k} \bm f_{k'} \beta_{k',i} \big)\big]$, and set $\bm f_k = \bm B \bm \psi_k$ for $\bm B = (\bm b(\tau_1), \ldots, \bm b(\tau_m))'$. In addition, we enforce the matrix orthonormality constraint $\bm F'\bm F = \utwi I_K$ on the basis matrix $\bm F = (\bm f_1,\ldots, \bm f_K)$: for each $k$, we (i) condition on the (linear) orthogonality constraints $\bm f_k ' \bm f_\ell = 0$ for $\ell \ne k$ and (ii) rescale $\bm f_k$ to unit norm. Notably,  $\bm F'\bm F = \bm I_K$ is satisfied for \emph{every} MCMC iteration. This constraint, coupled with the ordered MGP prior in Section \ref{shrinkage}, provides identifiability. 


The matrix orthonormality constraint $\bm F'\bm F = \bm I_K$ also provides important simplifications of the challenging likelihood in \eqref{reg}, which we leverage to achieve substantial improvements in computational efficiency for sampling the parameters in \eqref{reg}-\eqref{priors}. These computational gains are essential for moderate to large $p$, and produce algorithms for fully Bayesian inference that are empirically faster than existing alternatives. Consider the following:
\begin{lemma}\label{fdaLike2}
Under the constraint $\bm F' \bm F = \bm I_K$, the joint likelihood in \eqref{fdaVec} for $\{\beta_{k,i}\}$ is equivalent to the \emph{working likelihood} implied by
\begin{equation}\label{fdaVec2}
 y_{k,i} =  \beta_{k,i} + e_{k,i},
\quad e_{k,i} \stackrel{indep}{\sim} N(0, \sigma_{\epsilon}^2)
\end{equation}
up to a constant that does not depend on $\beta_{k,i}$, where $ y_{k,i} = \bm f_k' \bm Y_i $ and $e_{k,i} = \bm f_k' \bm \epsilon_i$.
\end{lemma}
The utility of Lemma \ref{fdaLike2} is that, for sampling the parameters in \eqref{reg}-\eqref{priors} which depend on $\beta_{k,i}$, we may replace the likelihood \eqref{fdaVec} with the simpler \eqref{fdaVec2}. Notably, \eqref{fdaVec2} depends on $m$ only via the projection $ y_{k,i} = \bm f_k' \bm Y_i $ and eliminates correlations among the components of $\bm Y_i$.

We sample the regression parameters $\{\mu_k, \alpha_{j,k}, \gamma_{k,i}\}_{j,k,i}$ jointly, which improves MCMC efficiency, and is a common strategy in Bayesian mixed effects models. This is accomplished in two steps: (i)  sample the regression coefficients $\{\mu_k, \alpha_{j,k}\}$ after marginalizing over the subject-specific effects $\{\gamma_{k,i}\}$, and (ii)  sample  $\{\gamma_{k,i}\}$ conditional on $\{\mu_k, \alpha_{j,k}\}$. The full conditional distributions are (i) $\left[\bm \alpha_k  | \bm Y, \cdots\right]\sim N\left(\bm Q_{\alpha_k}^{-1} \bm\ell_{\alpha_k}, \bm Q_{\alpha_k}^{-1}\right)$ for $\bm \alpha_k = (\mu_k, \alpha_{1,k},\ldots,\alpha_{p,k})'$, where $\bm Q_{\alpha_k} = \bm X' \bm \Sigma_{y_k}^{-1} \bm X + \bm \Sigma_{\alpha_k}^{-1}$ and $\bm \ell_{\alpha_k} = \bm X' \bm \Sigma_{y_k}^{-1} \bm{y}_k$ for $n \times (p + 1)$ design matrix $\bm X$, marginal variance $\bm \Sigma_{y_k} = \mbox{diag}\left(\{\sigma_{\gamma_{k,i}}^2 + \sigma_\epsilon^2\}_{i=1}^n\right)$,  prior variance $\bm \Sigma_{\alpha_k} = \mbox{diag}\left(\sigma_{\mu_k}^2, \sigma_{\alpha_{1,k}}^2,\ldots,\sigma_{\alpha_{p,k}}^2\right)$, and projected data $\bm{y}_k = (y_{k,1}, \ldots, y_{k,n})'$; and (ii) $\left[ \gamma_{k,i}  | \bm Y, \{\mu_k, \alpha_{j,k}\},\cdots\right] \stackrel{indep}{\sim} N\left( Q_{\gamma_{k,i}}^{-1} \ell_{\gamma_{k,i}},  Q_{\gamma_{k,i}}^{-1}\right)$, where   $Q_{\gamma_{k,i}} =  \sigma_\epsilon^{-2} + \sigma_{\gamma_{k,i}}^{-2}$ and $\ell_{\gamma_{k,i}} = \sigma_\epsilon^{-2}\left(y_{k,i} - \mu_{k} - \sum_{j=1}^p x_{i,j}  \alpha_{j,k}\right)$. A key feature of the proposed sampling strategy is that both component samplers (i) and (ii) are efficient. For the regression sampler (i), we apply \cite{rue2001fast} when $p < n$ with computational complexity $\mathcal{O}(p^3)$ and  \cite{bhattacharya2016fast} when $p>n$ with computational complexity $\mathcal{O}(n^2 p)$. The \cite{bhattacharya2016fast} sampler is designed for high dimensional Bayesian regression, but to the best of our knowledge has not been incorporated into Bayesian functional regression. The subject-specific effects sampler (ii) has computational complexity $\mathcal{O}(nK)$ for all $\{\gamma_{k,i}\}$, which allows us to incorporate subject-specific functional effects, modeled nonparametrically  in $\tau$ via $\{f_k\}$,  with minimal additional computational cost. In totality, the algorithm scales linearly in either $p$ or $n$.

The remaining sampling steps for the variance components in \eqref{fda} and \eqref{priors} consist of standard conjugate updates and parameter expansions, and are provided in the Appendix.

In Figure \ref{fig:compute}, we present empirical computing times for the proposed algorithm (*FOSR). For comparison with existing frequentist and Bayesian methods for FOSR, we include \cite{reiss2010fast}, which uses a generalized least squares estimation procedure (refund:GLS),  and \cite{goldsmith2016assessing}, which constructs a Gibbs sampler for a Bayesian FOSR model (refund:Gibbs), both implemented in the \texttt{refund} package in \texttt{R} \citep{refund}. The \texttt{refund} implementations require $p < n$. Figure \ref{fig:compute} demonstrates that the proposed algorithm is fast,  scalable, and superior to Bayesian and frequentist alternatives, and empirically validates linear time complexity in either $p$ or $n$. 
 A group lasso alternative \citep{barber2017function} was also considered, but omitted from Figure \ref{fig:compute} since it was noncompetitive.


\begin{figure}[h]
\begin{center}
\includegraphics[width=.49\textwidth]{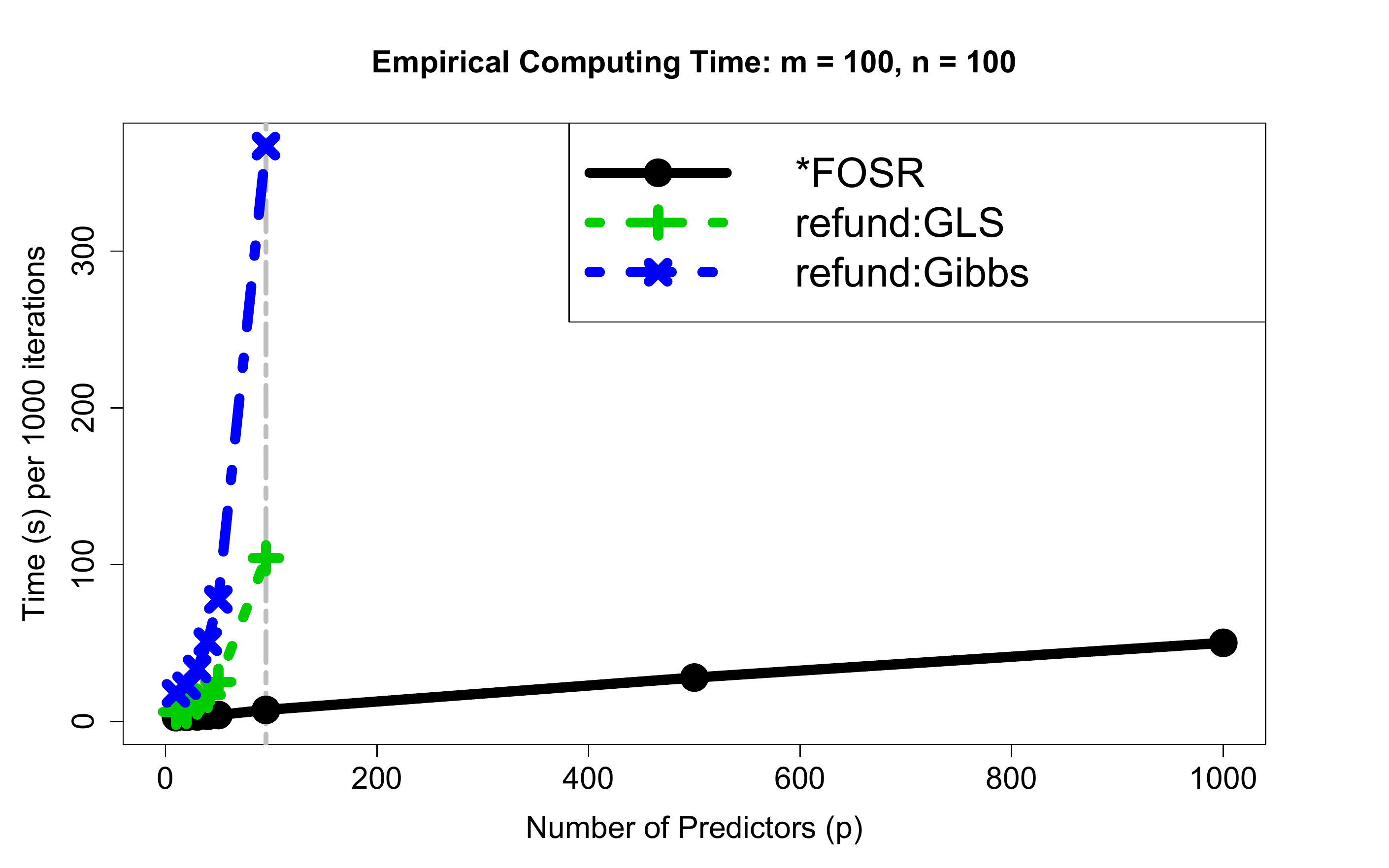}
\includegraphics[width=.49\textwidth]{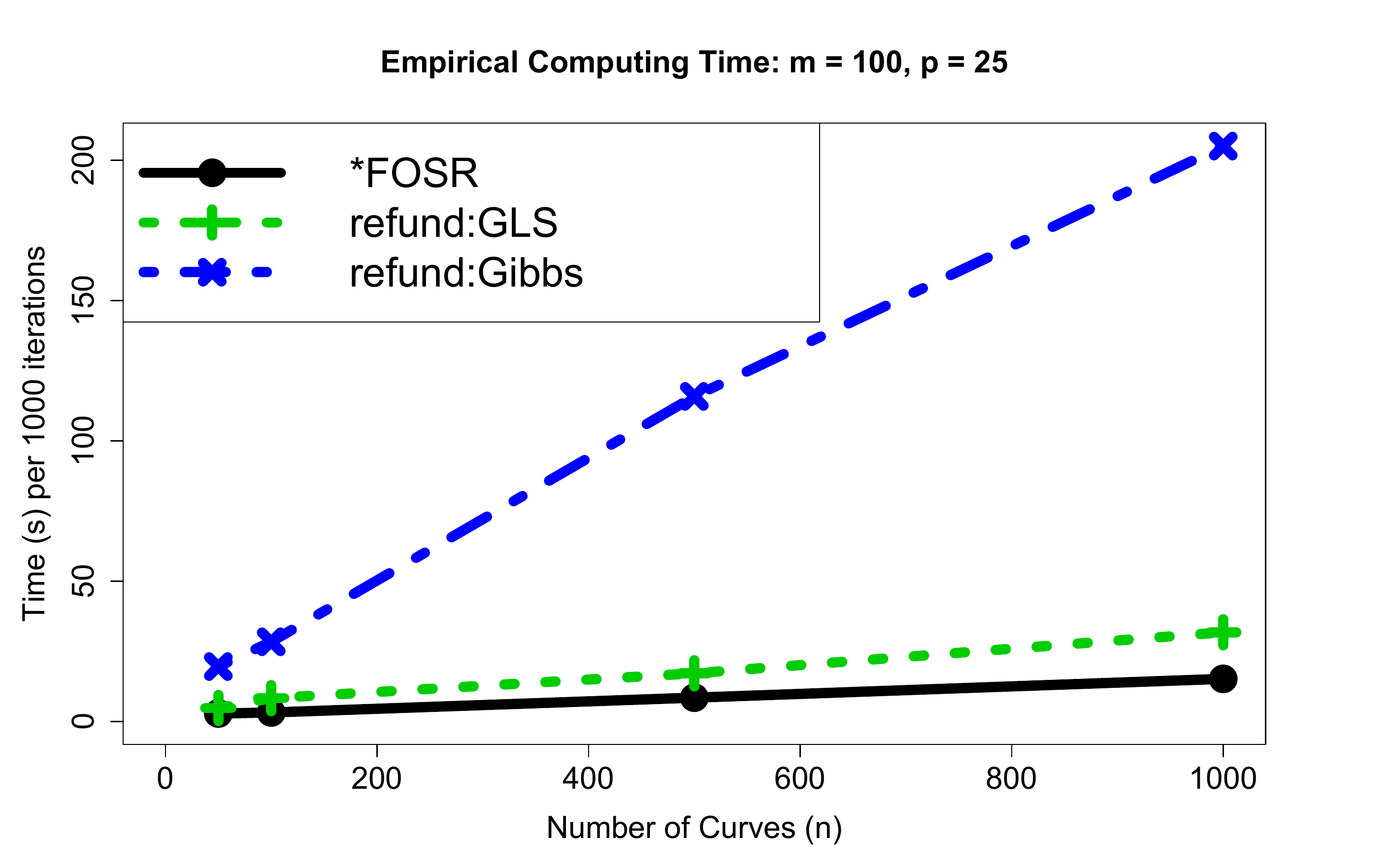}
\caption{Empirical computing time for several FOSR algorithms with varying $p$ {\bf (left)} and varying $n$ {\bf (right)} reported in seconds per 1000 MCMC iterations for Bayesian methods and total computing time for frequentist methods (using \texttt{R} on a MacBook Pro, 2.7 GHz Intel
Core i5). Each computing time is an average of 10 replicates. The vertical gray dashed line (left) indicates $p=95$, for which *FOSR runs in about 7 seconds. The proposed algorithm is empirically faster, with linear time complexity in either $p$ or $n$.
 \label{fig:compute}}
\end{center}
\end{figure}

\section{Variable Selection in Functional Regression}\label{dss}
Variable selection in FOSR is important for obtaining parsimonious and interpretable model summaries and reducing storage costs. Fundamentally, we are interested in identifying a subset of predictors that maintains nearly the predictive ability of the full model. The tradeoff between accuracy and sparsity may be defined precisely by a loss function; estimation and variable selection are achieved by minimizing the expected loss under the posterior predictive distribution. This approach, recently pioneered by \cite{hahn2015decoupling} for multiple linear regression, effectively \emph{decouples shrinkage and selection} (DSS):  posterior summarization, which may incorporate sparsity constraints for variable selection, is  distinct from the choice of the prior. Importantly, unlike \emph{marginal} selection approaches, such as the median probability model under sparsity priors   \citep{barbieri2004optimal}, hard-thresholding under shrinkage priors \citep{carvalho2010horseshoe}, or GBPVs \citep{meyer2015bayesian}, DSS selects variables \emph{jointly}, while accounting for collinearity among predictors.

Consider predicting new observations $\tilde Y_i(\tau_\ell)$ for each subject $i=1,\ldots,n$ for a pre-defined set of points $\{\tau_\ell\}_{\ell=1}^m$ and design points $\bm{\tilde x}_i = (\tilde x_{i,1},\ldots,\tilde x_{i,p})'$, which may differ from $\{x_{i,j}\}_{j=1}^p$.  Variable selection is provided by group sparsity: removing the $j$th predictor is equivalent to setting the target regression function $\tilde\delta_j(\tau) = 0$ for \emph{all} $\tau \in \mathcal{T}$. We write  $\tilde \delta_j(\cdot)$ evaluated at $\{\tau_\ell\}_{\ell=1}^m$ as follows: $\bm{\tilde\delta}_j = (\tilde \delta_j(\tau_1),\ldots, \tilde \delta_j(\tau_m))' = \bm F \bm \delta_j$, where $\bm F$ is an $m\times K$ basis matrix with $\bm F' \bm F = \bm I_K$ and $\bm \delta_j = (\delta_{j,1},\ldots,\delta_{j,K})'$ are the basis coefficients. For now, suppose that $\bm F$ is known: important examples include the identity matrix $\bm F = \bm I$, so that $\bm{\tilde\delta}_j = \bm{\delta}_j$, and the (orthonormal) spline matrix $\bm F = \bm B$; the case of unknown $\bm F$ from Section \ref{loadings} is considered subsequently. We propose the following loss function:
\begin{equation}\label{loss0}
\mathcal{L}(\bm{\tilde Y}, \bm{\Delta}) = \frac{1}{nm}\sum_{i=1}^n \big|\big| \bm{\tilde Y}_i - \bm F \bm{\delta}_0 - \bm F \bm{\Delta}\bm{\tilde x}_i\big|\big|_2^2 + \lambda \big|\big|\bm{\Delta}\big|\big|_0
\end{equation}
where $||\cdot||_2$ is the Euclidean norm,  $\bm{\tilde Y}_i = (\tilde Y_i(\tau_1),\ldots,\tilde Y_i(\tau_m))'$, $\bm{\delta}_0 = (\delta_{0,1},\ldots,\delta_{0, K})$ is the vector of basis coefficients for the functional intercept, $\bm{\Delta}$ is the $K\times p$ matrix with entries $(\bm{\Delta})_{k, j} = \delta_{j,k}$ the $k$th basis coefficient for predictor $j$,  and $ ||\bm{\Delta}||_0 = \sum_{j=1}^p \mathbb{I}\{\delta_{j,k} \ne 0 \mbox{ for some } k = 1,\ldots,K\}$. The predictive ability of $\{\tilde\delta_j(\cdot)\}$ is measured by the squared prediction error over all points $\tau_1,\ldots,\tau_m$ and subjects $i=1,\ldots,n$, while the second term in \eqref{loss0} encourages column-wise sparsity of the basis coefficients $\bm{\Delta}$ with the tradeoff determined by $\lambda > 0$. Since $\bm F' \bm F = \bm I_K$, sparsity of the basis coefficients $\bm \delta_j = \bm 0_K$ is equivalent to sparsity of the regression functions $\bm{\tilde\delta}_j = \bm 0_M$  evaluated at $\{\tau_\ell\}_{\ell=1}^m$. When $K \ll m$, the sparsity penalty in \eqref{loss0} operates in a lower-dimensional space, which produces more stable and efficient computations. 

To select predictors, we minimize the expectation of the loss function \eqref{loss0} under the posterior predictive distribution, $[\bm{\tilde Y} | \bm Y]$, which marginalizes over model parameters, say $\bm \theta$. Although  $[\bm{\tilde Y} | \bm Y]$ is unavailable in closed form, the conditional predictive distribution $[\bm{\tilde Y}  | \bm\theta]$ and the posterior distribution of the model parameters $[\bm \theta | \bm Y]$ are sufficient for obtaining a useful representation of the posterior predictive expected loss. The \emph{conditional} predictive distribution for $\bm{\tilde Y}_i$, after marginalizing over the subject-specific effects $\tilde\gamma_i(\cdot)$, is
 $[\bm{\tilde Y}_i  | \bm\theta]\stackrel{indep}{\sim} N(\bm F \bm{\mu}+ \bm F\utwi{A} \bm{\tilde x}_i, \bm{\Sigma}_i)$,
where $\bm \theta = \left(\{\mu_k\}, \{\alpha_{j,k}\}, \{\sigma_{\gamma_{k,i}}\}, \{f_k\},\sigma_\epsilon\right)$ are the relevant model parameters, $\bm{\mu} = (\mu_1,\ldots,\mu_K)'$, $\utwi{ A}$ is the $K\times p$ matrix with entries $(\utwi{A})_{k, j} = \alpha_{j,k}$, and $\bm{\Sigma}_i = \sigma_\epsilon^2 \bm I_m + \bm F \bm \Sigma_{\gamma_i} \bm F'$ with $\bm \Sigma_{\gamma_i} = \mbox{diag}\left(\{\sigma_{\gamma_{k,i}}^2\}_{k=1}^K\right)$. The expected loss may be simplified:


\begin{theorem}\label{cor-dss}
For a \emph{known} basis matrix $\bm F$ with $\bm F' \bm F = \bm I_K$, the posterior predictive expectation of the loss in \eqref{loss0} is
\begin{equation}\label{ppel0}
\mathbb{E}_{[\bm{\tilde Y} | \bm Y]}\mathcal{L}(\bm{\tilde Y}, \bm{\Delta}) = C(\bm Y) + \frac{1}{nm}\sum_{i=1}^n \big|\big|\left(\bm{\bar \mu} + \bm{\bar{A}}\bm{\tilde x}_i\right) - \left(\bm{\delta}_0 + \bm{\Delta}\bm{\tilde x}_i\right)\big|\big|_2^2 + \lambda ||\bm{\Delta}||_0
\end{equation}
where $C(\bm Y)$ is a constant that does not depend on $\bm{ \delta}_0$ and $\bm{\Delta}$, $\utwi{\bar{ \mu}} = \mathbb{E}[\bm{ \mu} | \bm Y]$, and $\utwi{\bar{ A}} = \mathbb{E}[\bm{{A}} | \bm Y]$.
\end{theorem}
The posterior predictive expected loss \eqref{ppel0} is a penalized regression with response $\utwi{\bar{\mu}} + \utwi{\bar{A}}\bm{\tilde x}_i$ constructed from posterior expectations  and a sparsity penalty on $\bm{\Delta}$. Notably, \eqref{ppel0} is valid under \emph{any} prior for $\bm \mu$ and $\bm A$ as long as the posterior means $\utwi{\bar{ \mu}}$ and $\utwi{\bar{ A}}$ exist, and therefore is broadly applicable for Bayesian FOSR models. For computational tractability, we replace the penalty $||\bm{\Delta}||_0$ with a convex relaxation: $||\bm{\Delta}||_1 = \sum_{j=1}^p \sqrt{K} ||\bm{\Delta}_j||_2$ for $\bm{\Delta}_j = (\delta_{j,1}, \ldots, \delta_{j, K})'$, which is the group lasso penalty \citep{yuan2006model}. To avoid overshrinkage, we use an adaptive group lasso penalty:
\begin{equation}\label{adapt}
||\bm{\Delta}||_{1^*} = \sum_{j=1}^p w_j ||\bm{\Delta}_j||_2, \quad w_j = 1/||\bm{\bar{A}}_{j}||_2
\end{equation}
for $\bm{\bar{A}}_{j}$ the posterior mean of $(\alpha_{j,1},\ldots,\alpha_{j,K})'$. The use of the adaptive group lasso is analogous to the use of the adaptive lasso in \cite{hahn2015decoupling}. The solution path for minimization of \eqref{ppel0} with modified penalty \eqref{adapt} may be computed using existing software, such as the \texttt{gglasso} package in \texttt{R} \citep{gglasso}.

When $\bm F$ is unknown, as in Section \ref{loadings}, the loss function  \eqref{loss0} depends on $\bm{\tilde\delta}_j = \bm F \bm \delta_j$, and therefore depends on unknown model parameters $\{f_k\}$. To account for the uncertainty in $\{f_k\}$, we  construct an \emph{expected} loss function directly by integrating over $\bm F$:
\begin{equation}\label{e-loss0}
\mathcal{EL}(\bm \Delta) = \mathbb{E}_{[\bm F | \bm Y]}\left\{ \mathbb{E}_{[\bm{\tilde Y} | \bm F, \bm Y]} \left[ \frac{1}{nm} \sum_{i=1}^n \big| \big| \bm{\tilde Y}_i -  \bm F \bm \delta_0 - \bm F \bm \Delta\bm{\tilde x}_i \big| \big|_2^2 + \lambda ||\bm{\Delta}||_0
\right]\right\}
\end{equation}
The additional complexity in \eqref{e-loss0} is due to the uncertainty of $\{f_k\}$, which requires successive averaging over the distributions $[\bm{\tilde Y} | \bm F, \bm Y]$ and $[\bm F | \bm Y]$ to obtain the posterior predictive expectation under  $[\bm{\tilde Y} | \bm Y]$. Note that in general, $[\bm{\tilde Y} | \bm F, \bm Y]$ is not equal to $ [\bm{\tilde Y} | \bm F]$: while it is often assumed that $\bm{\tilde Y}$ is conditionally independent of $\bm Y$ given \emph{all} model parameters, this distribution is only conditional on $\bm F$. We again obtain a convenient simplification:
\begin{theorem}\label{thm-rr-dss}
For an \emph{unknown} basis matrix $\bm F$ with $\bm F' \bm F = \bm I_K$, the expected loss $\mathcal{EL}(\bm \Delta)$ in \eqref{e-loss0} is equivalent to \eqref{ppel0}
up to constants that do not depend on $\bm{\delta}_0$ and $\bm{\Delta}$.
\end{theorem}
Despite the additional complexity introduced by modeling $\bm F$ as unknown in Section \ref{loadings}, the orthonormality constraint $\bm F'\bm F = \bm I_K$ provides a mechanism by which the expected loss \eqref{e-loss0} collapses into the penalized regression \eqref{ppel0}. As a result, optimization may proceed as before. Given a solution $\bm{\hat{\Delta}}$ to \eqref{ppel0} using the convex relaxation \eqref{adapt}, the proposed DSS estimator of the regression functions is $\mathbb{E}[\bm F \bm{\hat{\Delta}} | \bm Y] = \utwi{\bar F} \bm{\hat{\Delta}}$ where $\utwi{\bar F} = \mathbb{E}[\bm F | \bm Y]$.

We select the tuning parameter $\lambda > 0$ in \eqref{ppel0} by identifying a sparsified model for which the predictive ability is within a range of uncertainty of that of the full model. Consider the following notion of proportion of variability explained, analogous to $R^2$ in linear regression:
\begin{equation}\label{var-exp}
\rho^2 = \frac{\big|\big| \bm A \bm{\tilde X}'\big|\big|_F^2}{\mathbb{E}_{[\bm{\tilde Y} | \bm \theta]}\big|\big| \bm{\tilde Y}\big|\big|_F^2}=  \frac{\big|\big| \bm A \bm{\tilde X}'\big|\big|_F^2}{\big|\big| \bm A \bm{\tilde X}'\big|\big|_F^2+ \sum_{i=1}^n \mbox{tr}(\bm\Sigma_i)}
\end{equation}
where $\bm{\tilde X}$ is the $n \times p$ matrix of predictors $[\tilde x_{i,j}]_{i,j}$, $\bm{\tilde Y}$ is the $n \times m$ matrix of predictive values $[\tilde Y_i(\tau_\ell)]_{i,\ell}$,  $||\cdot||_F$ denotes the Frobenius norm, and $\mbox{tr}(\cdot)$ is the trace operator. Since $\sum_{i=1}^n \mbox{tr}(\bm\Sigma_i) = nm \sigma_\epsilon^2 + \sum_{i=1}^n \sum_{k=1}^K \sigma_{\gamma_{k,i}}^2$, the total variance in the denominator of \eqref{var-exp} includes the regression model term $\bm A \bm{\tilde X}'$, the observation error variance $\sigma_\epsilon^2$, and the subject-specific variances $\sigma_{\gamma_{k,i}}^2$. We exclude the intercept from \eqref{var-exp}, but this is not strictly necessary. 

For a given $\lambda$, let $\bm{\hat{\Delta}}_\lambda$ denote the solution to \eqref{ppel0} with the adaptive group lasso penalty. Model discrepancy, due to re-estimation under sparsity, contributes to total variance:
\begin{equation}\label{var-exp-sparse}
\rho_\lambda^2 = \frac{\big|\big| \bm A \bm{\tilde X}'\big|\big|_F^2}{\mathbb{E}_{[\bm{\tilde Y} | \bm \theta]}\big|\big| \bm{\tilde Y}\big|\big|_F^2 +  \big|\big| \bm A \bm{\tilde X}' - \bm{\hat{\Delta}}_\lambda \bm{\tilde X}'\big|\big|_F^2}
\end{equation}
which we may simplify as above using $\mathbb{E}_{[\bm{\tilde Y} | \bm \theta]}\big|\big| \bm{\tilde Y}\big|\big|_F^2 = \big|\big| \bm A \bm{\tilde X}'\big|\big|_F^2+ nm \sigma_\epsilon^2 + \sum_{i=1}^n \sum_{k=1}^K \sigma_{\gamma_{k,i}}^2$. 

We compare the posterior distributions of $\rho^2$ and $\rho_\lambda^2$ to assess disparities in predictive ability among sparsified models. We construct a posterior selection summary plot as follows: (i) using the \texttt{gglasso} package \citep{gglasso}, minimize \eqref{ppel0} with the penalty \eqref{adapt} for $\bm{\hat{\Delta}}_\lambda$ on a grid of $\lambda$ values; (ii) for each $\bm{\hat{\Delta}}_\lambda$, compute the posterior distribution of $\rho_\lambda^2$ by substituting the posterior draws of $\bm A$, $\sigma_\epsilon$, and $\sigma_{\gamma_{k,i}}$ into \eqref{var-exp-sparse}; and (iii) plot the expected value and 90\% credible intervals of $\rho_\lambda^2$ against model size $||\bm{\hat{\Delta}}_\lambda ||_0$. The resulting plot shows how predictive ability declines with sparsity, but importantly includes the accompanying uncertainty (see Figure \ref{fig:mesa_dss}). As a general guideline for model selection, we choose the smallest model for which the 90\% posterior credible interval for $\rho_\lambda^2$ contains $\mathbb{E}[\rho^2 | \bm Y]$.


\section{Simulations}\label{simulations}

\subsection{Simulation Design}
A simulation study was conducted to (i) compare the estimation accuracy of the proposed method against existing alternatives, (ii) evaluate uncertainty quantification among Bayesian FOSR models, and (iii) study the variable selection properties of the proposed DSS  procedure. We are primarily interested in how these properties vary in the number of predictors, so we consider $p \in \{20, 50, 500\}$ with fixed $p_1 = 10$ non-null predictors. All simulations use $n=100$ curves with $m =30$ equally-spaced points in $[0,1]$.


For each subject $i$, we simulate correlated predictors $\{x_{i,j}\}_{j=1}^p$ from a normal distribution with mean zero and covariance $\mathrm{Cov}(x_{i,j}, x_{i,j'}) = 0.75^{|j - j'|}$, with $p_1 = 10$ non-null predictors evenly-spaced from $1,\ldots,p$.  Functional observations are simulated based on model \eqref{fda}-\eqref{reg}. For the true basis functions, we set $f_1^*(\tau ) = 1/\sqrt{m}$ and for $k = 2, ..., K^* = 4$, we let $f_k^*$ be an orthogonal polynomial of degree $k$. For each non-null predictor $j$, we uniformly sample $K_j^*$ factors to be nonzero, where $K_j^*$ follows a Poisson$(1)$ distribution truncated to $[1, K^*]$, and draw the nonzero factor coefficients $\alpha_{j,k}^* \stackrel{indep}\sim N(0, 1/k^2)$. Each non-null predictor $j$ may be associated with the functional response via a subset of $\{f_k^*\}_{k=1}^{K^*}$. We simulate the true factors $\beta_{k,i}^*=\mu_k^* + \sum_{j=1}^{p}{x_{j,i} \alpha_{j,k}^*} + \gamma_{k,i}^*$, where  $\mu_k^* = 1/k$ and $\gamma_{k,i}^* \stackrel{indep}{\sim} N(0, 1/k^2)$, which incorporates subject-specific random effects. Based on the true curves  $Y_i^*(\tau) = \sum_{k=1}^{K^*}f_k^*(\tau)\beta_{k,i}^*$, the functional observations are $Y_i(\tau) = Y_i^*(\tau) + \sigma^* \epsilon_i^*(\tau)$, where $\epsilon_i^*(\tau) \stackrel{iid}{\sim} N(0, 1)$. The observation error standard deviation  $\sigma^*$ is determined by the \emph{root-signal-to-noise ratio} (RSNR): $\sigma^* =  \sqrt{\frac{\sum_{i=1}^n \sum_{j=1}^m (Y_i^*( \tau_j) - \bar Y^*)^2}{nm - 1}}\Big/\mbox{RSNR}$ where $\bar Y^*$ is the sample mean of $\{Y_i^*( \tau_j)\}_{j,i}$. We select RNSR = 5 for moderately noisy functional data.

\subsection{Methods For Comparison}
We implement model \eqref{fda}-\eqref{priors} with $K = 6 > K^* = 4$ (*FOSR). Using this posterior distribution, we compute the sparse DSS estimates from Section \ref{dss} (*FOSR-DSS). For comparison with *FOSR-DSS, we also select variables marginally using GBPVs, which retain a variable $j$ if the simultaneous credible bands for $\{\tilde \alpha_j(\tau_\ell)\}_{\ell=1}^m$ exclude zero for some $\tau_\ell$.

We include two variations of \eqref{fda}-\eqref{priors} based on alternative models for $\{f_k\}$: Basis-FPCA, which estimates $\{f_k\}$ as FPCs using  \cite{xiao2013fast}
with the number of FPCs selected to explain 99\% of the variability in $\{Y_i(\tau_\ell) - \bar Y(\tau_\ell) \}_{\ell, i}$, and Basis-Spline, which uses an (orthonormalized) LR-TPS basis for $\{f_k\}$. Both Basis-FPCA and Basis-Spline use normal-inverse-gamma priors for \eqref{priors}. Importantly, Basis-FPCA and Basis-Spline provide baselines for assessing the potential gains in point estimation and uncertainty quantification associated with the proposed shrinkage priors and the model for $\{f_k\}$.  Basis-FPCA and Basis-Spline are implemented using the proposed Gibbs sampler by omitting the basis function sampling step, and rely on the computational results of Section \ref{algorithm} for scalability.

Lastly, we include three FOSR methods from the \texttt{refund} package in \texttt{R} (see Figure \ref{fig:compute}): estimation with a group lasso for variable selection (refund:Lasso; \citealp{barber2017function}), estimation using generalized least squares (refund:GLS; \citealp{reiss2010fast}), and a Bayesian model using FPCs to estimate the residual
covariance (refund:Gibbs; \citealp{goldsmith2016assessing}). For $p = 500 > n = 100$, the \texttt{refund} methods are not computationally feasible.


We compare methods using three metrics. For point estimation, we use the root mean square error of the regression coefficient
functions, $\mbox{RMSE} = \sqrt{\frac{1}{pm}\sum_{j=1}^p \sum_{l=1}^m \left[
  \tilde \alpha_{j}(\tau_\ell) - \tilde \alpha_{j}^*(\tau_\ell)\right]^2}$,
where $\tilde \alpha_j (\tau_\ell)$ is the estimated regression coefficient
for predictor $j$ and observation point $\tau_\ell$ and
$\tilde \alpha_{j}^*(\tau_\ell) = \sum_{k=1}^{K^*} f_k^*(\tau_\ell)\alpha_{j,k}^*$
is the true regression coefficient.
The Bayesian methods use the posterior expectation of
$\tilde \alpha_j (\tau_\ell)$ as the estimator. For uncertainty quantification among the Bayesian methods, we compute  the mean credible interval
width for all regression coefficient functions,
$\mbox{MCIW} = \frac{1}{pm}\sum_{j=1}^{p}\sum_{\ell=1}^{m}\left[
  \tilde \alpha_{j}^{(97.5)}(\tau_\ell) -
  \tilde \alpha_{j}^{(2.5) }(\tau_\ell)\right]$, where
$\tilde \alpha_{j}^{(q)}(\tau_\ell)$ is the $q$\% quantile of the posterior distribution
for $\tilde \alpha_{j}(\tau_\ell)$, along with the empirical coverage probability,
$\frac{1}{pm}\sum_{j=1}^{p}\sum_{\ell=1}^{m} \mathbb{I}\{
  \tilde \alpha_{j}^{(2.5)} (\tau_\ell) \leq
  \tilde \alpha_{j}^{*}   (\tau_\ell) \leq
  \tilde \alpha_{j}^{(97.5)}(\tau_\ell) \}$. The goal is to achieve the smallest MCIW with the correct nominal coverage (0.95). For comparing variable selection techniques, we compute receiver-operating characteristic (ROC) curves. ROC curves plot the true positive rate, or sensitivity, against the false positive rate, or $1-$ specificity, as the decision threshold varies, where a positive predictor corresponds to a non-zero regression function.  ROC curves further toward the upper left corner indicate superiority of the method.

\subsection{Simulation Results}
\begin{figure}
\begin{center}
\includegraphics[width=1\textwidth]{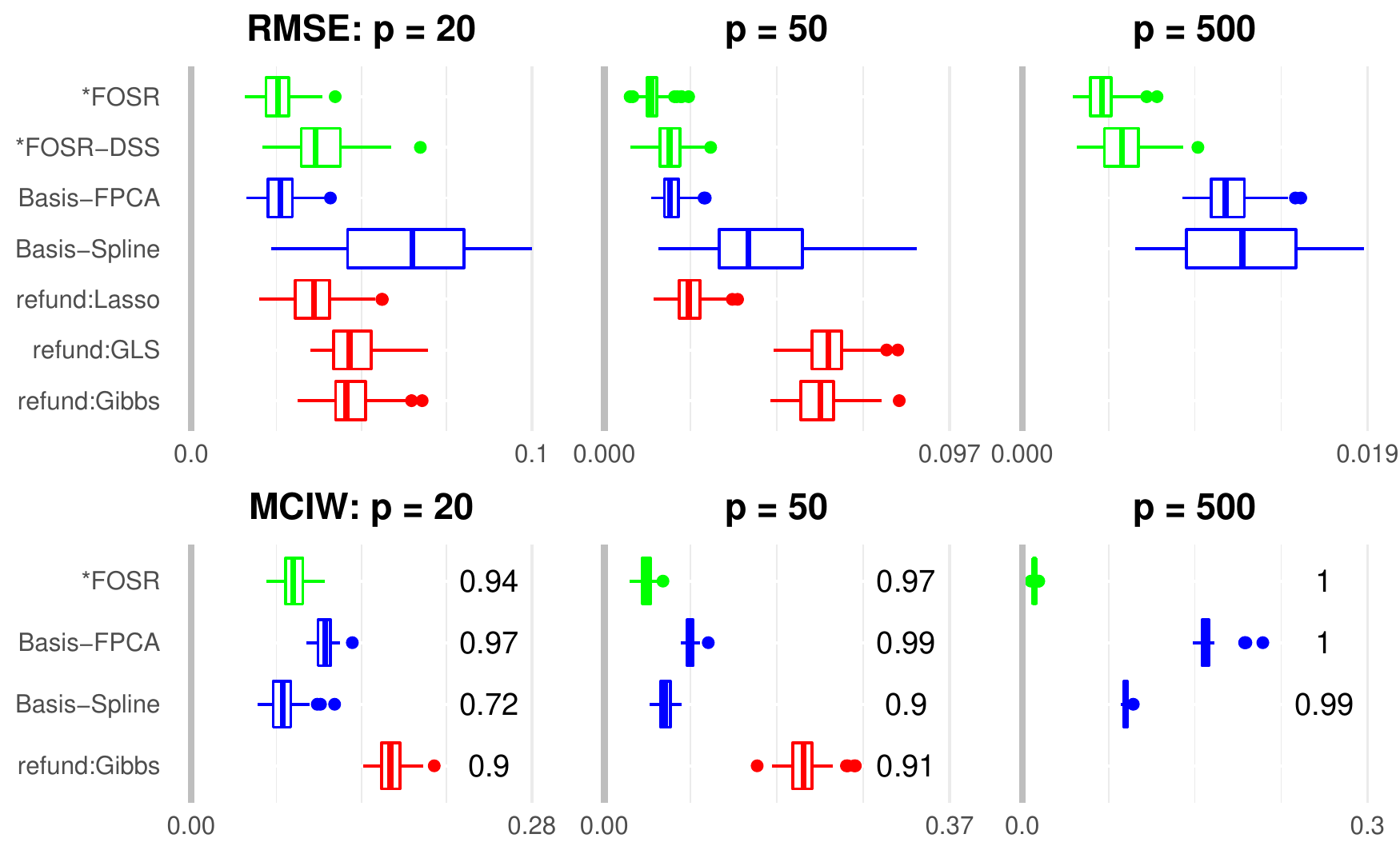}
\caption{
  Root mean squared errors \textbf{(top row)} and
  mean 95\% credible interval widths with empirical coverage probabilities \textbf{(bottom row)}
  for  $\tilde{\alpha}_j(\tau)$.
  The proposed methods (*FOSR and *FOSR-DSS) are in green, the $\{f_k\}$-modified methods (Basis-FPCA and Basis-Spline) are in blue, and existing methods (refund:Lasso, refund:GLS, and refund:Gibbs) are in red.
  Existing methods were not feasible for
$p = 500 > n = 100$.
  For display purposes, outliers of Basis-Spline are not shown.
\label{fig:sim_rmse}
}
\end{center}
\end{figure}

Figure \ref{fig:sim_rmse} displays RMSEs and MCIWs for 100 simulated datasets. The RMSEs show that *FOSR provides the best point estimation for all $p$, with the most substantial improvements for moderate to large $p \ge 50$.  The sparse estimates from *FOSR-DSS perform at least as well as refund:Lasso for $p=20$, and are outperformed only by *FOSR for $p \ge 50$. The case of $p=500$ demonstrates the utility of the shrinkage priors of Section \ref{shrinkage}, as Basis-FPCA and Basis-Spline are clearly dominated by *FOSR and *FOSR-DSS. Estimation of $\{f_k\}$ is also important: there is a sizable gap for all $p$ between Basis-Spline, which does \emph{not} estimate $\{f_k\}$, and *FOSR, *FOSR-DSS, and Basis-FPCA, which do estimate $\{f_k\}$.

The MCIWs in Figure \ref{fig:sim_rmse} demonstrate that the proposed *FOSR provides significantly narrower credible intervals than competing Bayesian methods, particularly for  $p \ge 50$, while maintaining approximately the correct nominal coverage. Note that although Basis-Spline produces narrow credible intervals, it suffers from severe undercoverage. The large improvements of *FOSR relative to Basis-FPCA suggest that the proposed shrinkage priors and model for the basis functions---which, unlike Basis-FPCA, accounts for the uncertainty of the unknown $\{f_k\}$---are both important for more precise uncertainty quantification.




\begin{figure}
\begin{center}
\includegraphics[width=1.0\textwidth]{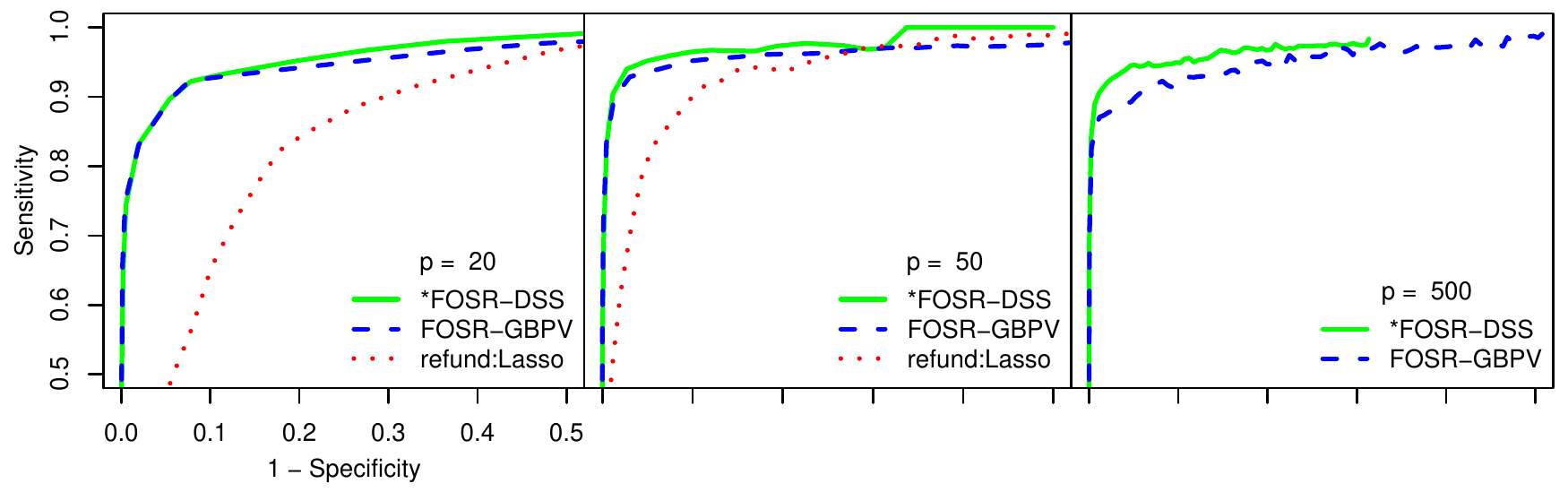}
\caption{
ROC curves to compare variable selection techniques.  Each point along the ROC curve is the average sensitivity and
  $1-$ specificity of a given model size. For $p=500$, *FOSR-DSS selects only a few false positives, so the corresponding ROC curve does not extend to lower (i.e., worse) specificity.
\label{fig:roc_curves}
}
\end{center}
\end{figure}

In Figure \ref{fig:roc_curves}, we show ROC curves for the competing variable selection techniques: *FOSR-DSS, FOSR-GBPV, and refund:Lasso.
For all $p$, the proposed *FOSR-DSS is at least as good as the other methods, with greater improvements relative to FOSR-GBPV as $p$ increases. Notably, refund:Lasso is inferior to both Bayesian methods for variable selection.




\section{Time-of-Day Physical Activity Levels for Elderly Adults}\label{app}
We analyze time-of-day physical activity levels for an elderly population obtained from
the National Sleep Research Resource \citep{dean2016scaling, zhang2018national}
in conjunction with the Multi-Ethnic Study of Atherosclerosis.
The dataset consists of actigraphy measurements and an accompanying sleep
questionnaire (see the Appendix) for black, white, Hispanic, and Chinese-American men and women. The aim of our investigation was to identify question items from a sleep questionnaire that are predictive of time-of-day physical activity levels, and to estimate and perform inference on these time-of-day effects.



To focus primarily on waking hours, we considered activity levels from 6am to 10pm on Wednesday and Saturdays. The raw count activity data was summed into 20 minute time increments and modeled as a function of time-of-day  with $m = 48$ observation points. Days with more than 10\% missingness typically had long periods of no activity recorded and were removed from the analysis. Six percent of the remaining days had missing observation points, which were imputed automatically within the Gibbs sampler. Covariate information  included age, gender, race/ethnicity, a Wednesday/Saturday indicator, and questionnaire responses. Due to substantial missingness among questionnaire responses, we modeled questionnaire items as categorical variables, grouped into high, low, and missing responses where appropriate (see the Appendix for details). In total, 2059 people ages 54-95 were considered over a cumulative $n=3568$ days, with 34 items from the questionnaire and $p = 74$.



Posterior samples from model \eqref{fda}-\eqref{priors} were obtained using the MCMC  algorithm in the Appendix. We report results for $K = 6$; larger values of $K$ produced similar results. Traceplots demonstrate good mixing and suggest convergence (see the Appendix).

\begin{figure}
\begin{center}
\includegraphics[width=0.6\textwidth]{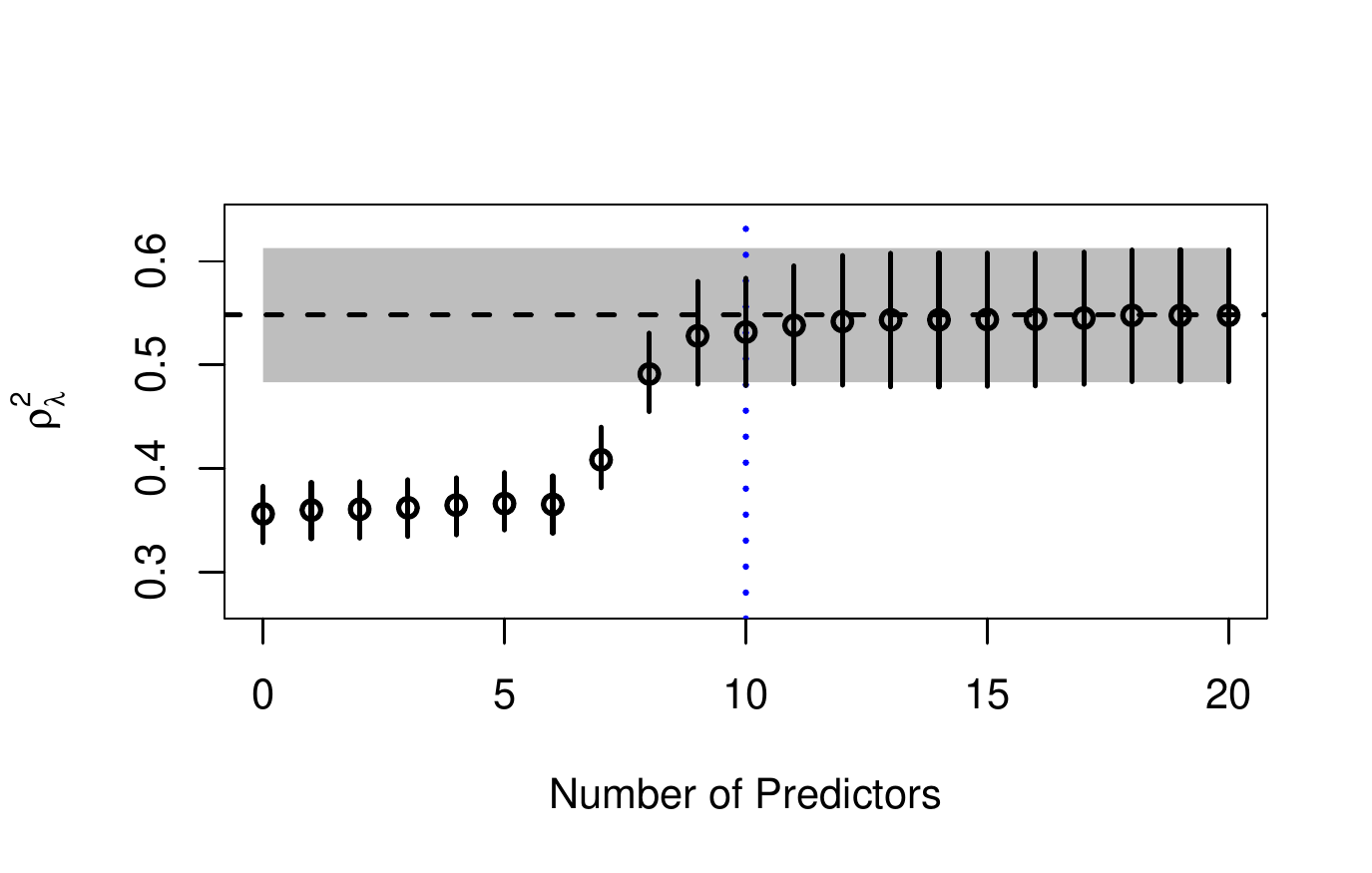}
\caption{
The selection summary plot for the time-of-day physical activity data.  Shown is the proportion of variability explained (with 95\% credible intervals) for
increasing model sizes.
The horizontal line denotes the proportion of variability explained by the full model with the gray band denoting the 95\% credible interval.
As model size increases, the explanatory power of the model increases. The vertical blue line denotes the selected model with ten predictors.
\label{fig:mesa_dss}
}
\end{center}
\end{figure}

We applied the DSS procedure from Section \ref{dss} to identify  demographic variables and questionnaire items that predict intraday physical activity. We perform selection on the \emph{category levels} rather than the questionnaire items: selected levels imply inclusion of the corresponding questionnaire item (along with the baseline category), but do not require other levels be included. The selection summary plot in Figure \ref{fig:mesa_dss} displays the tradeoff between predictive accuracy and sparsity: the proportion of variability explained increases quickly between models of size six and ten, but does not notably increase for larger models. The most reasonable model size is between eight and ten, while larger models do not offer additional predictive ability. Guided by Figure \ref{fig:mesa_dss}, we select ten predictors:
\texttt{Saturday}, \texttt{Gender}, \texttt{age}, and the seven categories of the four questionnaire items in Table \ref{table:selected}. Impressively, these four questionnaire items (along with \texttt{Saturday}, \texttt{Gender}, and \texttt{age}) retain nearly the predictive ability of the full questionnaire.

\begin{table}[h]
\begin{center}
\begin{tabular}{ p{1in} p{2.5in} p{2.5in} }
Item name & Categories & Question Content \\
\hline
  \texttt{bedtmwkday}    & missing/{\color{blue}\textbf{5-7}}/{\color{blue}\textbf{8-9}}/10-11/{\color{blue}\textbf{later}}         & Bedtime Weekday \\
  \texttt{nap}           & missing/{\color{blue}\textbf{low}}/high                                      & Usual Week: Number Of Naps \\
  \texttt{types}         & missing/{\color{blue}\textbf{evening}}/morning/neither                       & Type Of Person: Morning Or Evening \\
  \texttt{wkdaysleepdur} & missing/\textless7/{\color{blue}\textbf{7-9}}/{\color{blue}\textbf{\textgreater9}} & Sleep duration weekday \\
\end{tabular}
\caption{
  The questionnaire item name, categories, and question content
  of the items selected by DSS.
  The first category in each row is the contrasting variable and the selected
  categories are in blue and bolded.
  \label{table:selected}
}
\end{center}
\end{table}



\begin{figure}
\begin{center}
\includegraphics[width=1.0\textwidth]{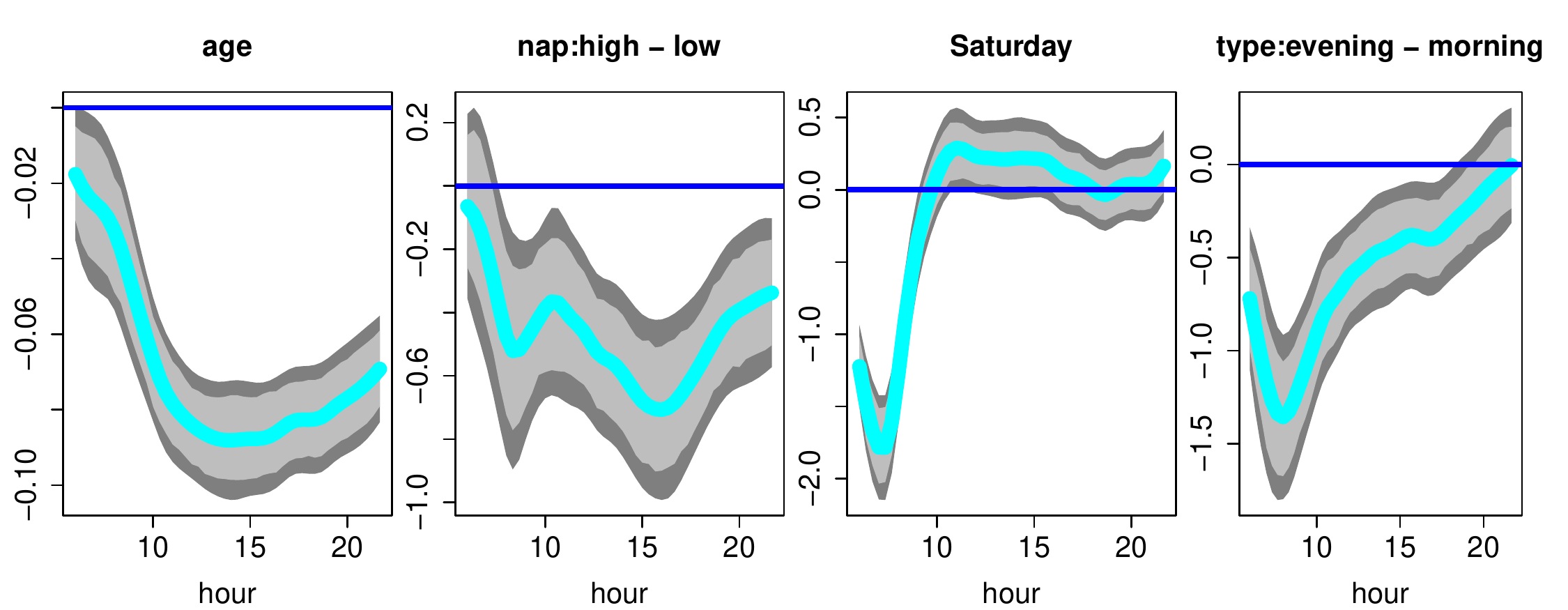}
\caption{
  Estimated regression coefficient functions for DSS-selected variables,
  95\% pointwise credible intervals (light gray), and 95\% simultaneous credible bands
  (dark grey).
  A horizontal blue line denotes zero change in activity. 
\label{fig:reg_coef}
}
\end{center}
\end{figure}

The estimated coefficient functions and 95\% simultaneous credible bands for several DSS-selected variables are displayed in Figure \ref{fig:reg_coef}. Most notably, these effects are time-varying, which confirms the importance of using a \emph{functional} regression model. The simultaneous credible bands for age are below zero for all times of day, indicating that daily activity declines with age. This effect is largest from 10am to 4pm, which we note is the most active time of day in the study. Individuals who reported frequent napping also tended to be less active throughout the day, particularly in the mid-afternoon. Physical activity levels were lower on Saturday mornings compared to Wednesday mornings, but were similar throughout the rest of the day. Lastly, self-described ``evening people" were less active in the morning than ``morning people", but more surprisingly, were also less active overall.


\section{Conclusions}\label{conclusions}
We developed a fully Bayesian framework for function-on-scalars regression with many predictors. Nonparametric and unknown basis functions were proposed for greater modeling flexibility and proper uncertainty quantification via the posterior distribution. Carefully-designed shrinkage priors were employed to minimize the impact of unimportant predictor variables, which is particularly important for moderate to large $p$. We introduced a novel variable selection technique for functional regression, which identifies a sparse subset of predictors that minimizes loss in predictive performance relative to the full model. A simulation study illustrated the improvements in point estimation, uncertainty quantification, and variable selection offered by the proposed methodology.  Full posterior inference was provided by an efficient Gibbs sampler with computational
complexity scaling in either $n$ or $p$. The methodology was applied to an actigraphy and sleep questionnaire dataset from the National Sleep Research Resource.

The proposed methodology offers several promising extensions. Modifications for binomial and count data are available by coupling the computational approach of Section \ref{algorithm} with well-known Gaussian parameter expansions. Alternative models for $\{f_k\}$ may be introduced for other applications, for example wavelets for non-smooth functional data or Fourier basis functions for periodic functional data. Lastly, the proposed DSS approach offers a general framework for Bayesian variable selection and posterior summarization in other functional, spatial, and time series regression problems.




\bibliographystyle{apalike}
\bibliography{refs}

\appendix \section*{Appendix}

\subsection*{MCMC Sampling Algorithm}
The MCMC sampling algorithm consists of the following steps:
\begin{enumerate}
\item {\bf Imputation:}   $\left[Y_i(\tau_i^*) | \cdots \right] \stackrel{indep}{\sim} N\big(\sum_k f_k(\tau_i^*) \beta_{k,i}, \sigma_{\epsilon}^2\big)$ for all unobserved $Y_i(\tau_i^*)$.

\item {\bf Loading curves and smoothing parameters:} for $k=1,\ldots,K$,
\begin{enumerate}
\item  $[\lambda_{f_k} | \cdots] \sim \mbox{Gamma}((L_m - D + 1 + 1)/2, \bm\psi_k' \bm\Omega \bm\psi_k/2)$ truncated to $(10^{-8}, \infty)$.
\item  $\left[\bm \psi_k | \cdots \right] \sim N\left(\bm Q_{\psi_k}^{-1} \bm \ell_{\psi_k}, \bm Q_{\psi_k}^{-1}\right)$ conditional on $\bm C_k \bm \psi_k = \bm 0$, where
$\bm Q_{\psi_k} =  \sigma_\epsilon^{-2}(\bm B'\bm B)\sum_{i=1}^n \beta_{k,i}^2 + \lambda_{f_k} \bm \Omega$, $\bm \ell_{\psi_k} =  \sigma_\epsilon^{-2}\bm B' \sum_{i=1}^n \big[\beta_{k,i} \big( \bm Y_i - \sum_{k' \ne k} \bm f_{k'} \beta_{k',i} \big)\big]
$, $\bm C_k = (\bm f_1, \ldots, \bm f_{k-1}, \bm f_{k+1}, \ldots, \bm f_K)' \bm B= (\bm \psi_1, \ldots, \bm \psi_{k-1}, \bm \psi_{k+1}, \ldots, \bm \psi_K)'$ using the following:
\begin{enumerate}
\item Compute the (lower triangular) Cholesky decomposition $\bm Q_{\psi_k}   = \bar{\bm{Q}}_L\bar{\bm{Q}}_L'$;
\item Use forward substitution to obtain $\bar{\bm{\ell}}$ as the solution to $\bar{\bm{Q}}_L \bar{\bm{\ell}} =\bm \ell_{\psi_k}$,  then use backward substitution to obtain $\bm \psi_k^0$ as the solution to $\bar{\bm{Q}}_L' \bm \psi_k^0 = \bar{\bm{\ell}} + {\bm{z}}$, where ${\bm{z}}\sim N (\bm{0},  \bm{I}_{L_m})$;
\item Use forward substitution to obtain $\utwi{\bar C}$ as the solution to $\bar{\bm{Q}}_L \utwi{\bar C}= \bm C_k $,  then use backward substitution to obtain $\utwi{\tilde C}$ as  the solution to $\bar{\bm{Q}}_L'\utwi{\tilde C} = \utwi{\bar C}$;
\item Set $\bm \psi_k^* = \bm \psi_k^0 - \utwi{\tilde C} (\bm C_k\utwi{\tilde C})^{-1}\bm C_k \bm \psi_k^0$;
\item Retain the vectors $\bm \psi_k = \bm \psi_k^*/\sqrt{{\bm \psi_k^*}'\bm B' \bm B\bm \psi_k^*} = \bm \psi_k^*/ || \bm \psi_k^*||$  and $\bm f_k = \bm B \bm \psi_k$ and update $\beta_{k,i} \leftarrow \beta_{k,i} || \bm \psi_k^*||$.
\end{enumerate}
\end{enumerate}

\item {\bf Project:} update $ y_{k,i} = \bm f_k' \bm Y_i = \bm \psi_k' \left(\bm B' \bm Y_i\right)$ for all $k,i$.

\item {\bf Regression coefficients and subject-specific effects}:  

\begin{enumerate}
\item $\left[\bm \alpha_k  | \bm Y, \cdots\right]\sim N\left(\bm Q_{\alpha_k}^{-1} \bm\ell_{\alpha_k}, \bm Q_{\alpha_k}^{-1}\right)$ for $k=1,\ldots,K$ using Algorithm \ref{pln} if $p < n$ or Algorithm \ref{pgn} if $p > n$, where $\bm Q_{\alpha_k} = \bm X' \bm \Sigma_{y_k}^{-1} \bm X + \bm \Sigma_{\alpha_k}^{-1}$ and $\bm \ell_{\alpha_k} = \bm X' \bm \Sigma_{y_k}^{-1} \bm{y}_k$ for $n \times (p + 1)$ design matrix $\bm X$, marginal variance $\bm \Sigma_{y_k} = \mbox{diag}\left(\{\sigma_{\gamma_{k,i}}^2 + \sigma_\epsilon^2\}_{i=1}^n\right)$,  prior variance $\bm \Sigma_{\alpha_k} = \mbox{diag}\left(\sigma_{\mu_k}^2, \sigma_{\alpha_{1,k}}^2,\ldots,\sigma_{\alpha_{p,k}}^2\right)$, and projected data $\bm{y}_k = (y_{k,1}, \ldots, y_{k,n})'$.

\item 
$\left[ \gamma_{k,i}  | \bm Y, \{\mu_k, \alpha_{j,k}\},\cdots\right] \stackrel{indep}{\sim} N\left( Q_{\gamma_{k,i}}^{-1} \ell_{\gamma_{k,i}},  Q_{\gamma_{k,i}}^{-1}\right)$ for $i=1,\ldots,n$ and $k=1,\ldots,K$, where   $Q_{\gamma_{k,i}} =  \sigma_\epsilon^{-2} + \sigma_{\gamma_{k,i}}^{-2}$ and $\ell_{\gamma_{k,i}} = \sigma_\epsilon^{-2}\left(y_{k,i} - \mu_{k} - \sum_{j=1}^p x_{i,j}  \alpha_{j,k}\right)$.
\end{enumerate}

\item {\bf Variance parameters:}
\begin{enumerate}

\item {\bf Observation error variance:} $[\sigma_\epsilon^{-2} | \cdots] \sim \mbox{Gamma}\left(\frac{mn}{2}, \frac{1}{2}\sum_{i=1}^n ||\bm Y_i - \bm F \bm\beta_i ||^2\right)$

\item {\bf Multiplicative Gamma Process Parameters:} given $\mu_k$ and $\gamma_{k,i}$,
\begin{enumerate}

\item  $[\delta_{\mu_1} | \cdots ] \sim \mbox{Gamma}\big(a_{\mu_1} + \frac{K}{2}, 1 + \frac{1}{2} \sum_{k=1}^K \tau_{\mu_k}^{(1)} \mu_k^2 \big)$ and $[\delta_{\mu_\ell} | \cdots ] \sim \mbox{Gamma}\big(a_{\mu_2} + \frac{K - \ell + 1}{2}, 1 + \frac{1}{2} \sum_{k=\ell}^K \tau_{\mu_k}^{(\ell)} \mu_k^2 \big)$ for $\ell > 1$ and $\tau_{\mu_\ell}^{(k)} = \prod_{h = 1, h \ne k}^\ell \delta_{\mu_h}$.

\item Set $\sigma_{\mu_k} = \prod_{\ell \le k} \delta_{\mu_\ell}^{-1/2}$.

\item  $[\delta_{\gamma_1} | \cdots ] \sim \mbox{Gamma}\big(a_{\gamma_1} + \frac{Kn}{2}, 1 + \frac{1}{2} \sum_{k=1}^K \tau_{\gamma_k}^{(1)} \sum_{i=1}^n\gamma_{k,i}^2 \xi_{\gamma_{k,i}} \big)$ and $[\delta_{\gamma_\ell} | \cdots ] \sim \mbox{Gamma}\big(a_{\gamma_2} + \frac{(K - \ell + 1)n}{2}, 1 + \frac{1}{2} \sum_{k=\ell}^K \tau_{\gamma_k}^{(\ell)} \sum_{i=1}^n \gamma_{k,i}^2 \xi_{\gamma_{k,i}} \big)$ for $\ell > 1$ and $\tau_{\gamma_\ell}^{(k)} = \prod_{h = 1, h \ne k}^\ell \delta_{\gamma_h}$.
\item Set $\sigma_{\gamma_k} = \prod_{\ell \le k} \delta_{\gamma_\ell}^{-1/2}$
\item  $[\xi_{\gamma_{k,i}} | \cdots] \stackrel{indep}{\sim} \mbox{Gamma}\big(
\frac{\nu_\gamma}{2} + \frac{1}{2}, \frac{\nu_\gamma}{2} + \frac{\gamma_{k,i}^2}{2\sigma_{\gamma_k}^2}
\big)$
\item Set $\sigma_{\gamma_{k,i}} =  \sigma_{\gamma_k}/\sqrt{\xi_{\gamma_{k,i}}}$.
\end{enumerate}

\item {\bf Hierarchical Half-Cauchy Parameters:} using parameter expansions for the half-Cauchy distribution \citep{wand2011mean}, 
\begin{enumerate}
\item  $[\sigma_{\alpha_{j,k}}^{-2} | \cdots ] \stackrel{indep}{\sim}  \mbox{Gamma}\big(1, \xi_{{\alpha_{j,k}}} + \alpha_{j,k}^2/2\big)$,
 $[\xi_{{\alpha_{j,k}}} | \cdots] \stackrel{indep}{\sim}  \mbox{Gamma}\big(1, \lambda_{j}^{-2} + \alpha_{\omega_{j,k}}^{-2}\big)$.


\item $[\lambda_{j}^{-2} | \cdots] \stackrel{indep}{\sim}  \mbox{Gamma}\big(\frac{K + 1}{2}, \xi_{\lambda_{j}} + \sum_{k=1}^K \xi_{\lambda_{j,k}} \big)$,
 $[\xi_{\lambda_{j}} | \cdots] \stackrel{indep}{\sim}  \mbox{Gamma}\big(1, \lambda_{0}^{-2} + \lambda_{j}^{-2}\big)$.

\item  $[\lambda_{0}^{-2} | \cdots] \sim \mbox{Gamma}\big(\frac{p + 1}{2}, \xi_{\lambda_{0}} + \sum_{j=1}^p \xi_{\lambda_{j}} \big)$,
$[\xi_{\lambda_{0}} | \cdots] \sim \mbox{Gamma}\big(1, p + \lambda_{0}^{-2} \big)$.
\end{enumerate}

\end{enumerate}
\item {\bf Hyperparameters:} sample $a_{\mu_1}, a_{\mu_2}, a_{\gamma_1},a_{\gamma_2}$, and $\nu_\gamma$ independently using the slice sampler \citep{neal2003slice}.
\end{enumerate}

The sampling algorithms for the regression coefficients are as follows:
\begin{algorithm}\label{pln}($p < n$)
\begin{enumerate}
\item Sample $\bm \delta \sim N(\bm 0, \bm I_n)$.
\item Compute the (lower triangular) Cholesky decomposition $\bm Q_{\alpha_k}   = \bar{\bm{Q}}_L\bar{\bm{Q}}_L'$.
\item Use forward substitution to obtain $\bar{\bm{\ell}}$ as the solution to $\bar{\bm{Q}}_L \bar{\bm{\ell}} =\bm \ell_{\alpha_k}$.
\item  Use backward substitution to obtain $\bm \alpha_k^*$ as the solution to $\bar{\bm{Q}}_L' \bm \alpha_k^* = \bar{\bm{\ell}} + \bm \delta$.
\end{enumerate}
\end{algorithm}
\begin{algorithm}\label{pgn}($p > n$)
\begin{enumerate}
\item Sample $\bm u \sim N(\bm 0, \bm \Sigma_{\alpha_k})$ and $\bm \delta \sim N(\bm 0, \bm I_n)$ independently.
\item Compute $\bm{X}_{k} = \bm \Sigma_{y_k}^{-1/2} \bm X$ for $\bm \Sigma_{y_k}^{-1/2} = \mbox{diag}( 1\big/\sqrt{\sigma_{\gamma_{k,i}}^2 + \sigma_\epsilon^2})$.
\item Set $\bm v = \bm{X}_k\bm u + \bm \delta$.
\item Solve $\left(\bm{X}_k \bm \Sigma_{\alpha_k} \bm{X}_k' + \bm I_n\right) \bm w = ( \bm \Sigma_{y_k}^{-1/2} \bm{y}_k  -  \bm v)$ to obtain $\bm w$.
\item Set $\bm \alpha_k^* = \bm u + \bm \Sigma_{\alpha_k} \bm{X}_k'\bm w$.
\end{enumerate}
\end{algorithm}

\subsection*{Theoretical Results}
\begin{proof}[{Proof (Lemma 1)}]
The joint likelihood under the model  $\bm Y_i = \sum_{k=1}^K \bm f_k \beta_{k,i} + \bm \epsilon_i$ with $\bm \epsilon_i \sim N(\bm 0, \sigma_{\epsilon}^2 \bm I_m)$ and subject to  $\bm F' \bm F = \bm I_K$ is 
$$
p\left(\bm Y_1,\ldots, \bm Y_n | \{\bm f_k\}, \{\beta_{k,i}\}, \sigma_{\epsilon}\right)
 = c_Y \sigma_\epsilon^{-mn} \prod_{i=1}^n \exp\left\{
 -\frac{1}{2 \sigma_{\epsilon}^2} \left[\bm Y_i ' \bm Y_i + \bm \beta_i' \bm\beta_i - 2 \bm \beta_i' \left(\bm F' \bm Y_i\right)
 \right]
 \right\}
$$
where $c_Y = (2\pi)^{-mn/2}$ is a constant and $\bm \beta_i' = (\beta_{1,i},\ldots,\beta_{K,i})$. By inspection, this likelihood is proportional to the likelihood implied by the model $ y_{k,i} =  \beta_{k,i} + e_{k,i}$ with $e_{k,i} \stackrel{indep}{\sim} N(0, \sigma_{\epsilon}^2)$ up to a constant that does not depend on $\{\beta_{k,i}\}$, where $ y_{k,i} = \bm f_k' \bm Y_i $ and $e_{k,i} = \bm f_k' \bm \epsilon_i$.
\end{proof}

\begin{proof}[{Proof (Theorem 1)}]
Ignoring the intercept $\bm F \bm \delta_0$ for simplicity, the posterior predictive expected loss is
\begin{align*}
\mathbb{E}_{[\bm{\tilde{Y}} | \bm Y]}\mathcal{L}(\bm{\tilde Y}, \bm{\Delta}) &= \mathbb{E}_{[\bm{\tilde{Y}} | \bm Y]}\Big[\frac{1}{nm}\sum_{i=1}^n \big|\big| \bm{\tilde Y}_i - \bm F \bm{\Delta}\bm{\tilde x}_i\big|\big|_2^2 + \lambda \big|\big|\bm{\Delta}\big|\big|_0\Big] \\
&= \frac{1}{nm}\sum_{i=1}^n \mathbb{E}_{[\bm{\tilde{Y}} | \bm Y]}\big|\big| \bm{\tilde Y}_i  - \bm F \bm{\Delta}\bm{\tilde x}_i\big|\big|_2^2 + \lambda \big|\big|\bm{\Delta}\big|\big|_0
\end{align*}
The expectation term may be written $\mathbb{E}_{[\bm{\tilde{Y}} | \bm Y]}\big|\big| \bm{\tilde Y}_i  - \bm F \bm{\Delta}\bm{\tilde x}_i\big|\big|_2^2 =  \mathbb{E}_{[\bm\theta | \bm Y]}\left[\mathbb{E}_{[\bm{\tilde Y} | \bm\theta ]}\big|\big| \bm{\tilde Y}_i  - \bm F \bm{\Delta}\bm{\tilde x}_i\big|\big|_2^2\right]$, which we compute in two steps. First,
\begin{align*}
\mathbb{E}_{[\bm{\tilde Y} | \bm\theta ]}\big|\big| \bm{\tilde Y}_i - \bm F \bm{\Delta}\bm{\tilde x}_i\big|\big|_2^2 &= \mathbb{E}_{[\bm{\tilde Y} | \bm\theta ]}\Big|\Big| \left[\bm{\tilde Y}_i - \bm{\hat Y}_i(\bm\theta)\right] + \left[\bm{\hat Y}_i(\bm\theta) -  \bm F \bm{\Delta}\bm{\tilde x}_i\right]\Big|\Big|_2^2\\
&= \hat v(\bm\theta) + \Big|\Big| \bm{\hat Y}_i(\bm\theta) -  \bm F \bm{\Delta}\bm{\tilde x}_i \Big|\Big|_2^2
\end{align*}
where $\hat v(\bm\theta) \equiv \mathbb{E}_{[\bm{\tilde Y} | \bm\theta]}\big|\big|\bm{\tilde Y}_i - \bm{\hat Y}_i(\bm\theta)\big|\big|_2^2 < \infty$ and $\bm{\hat Y}_i(\bm\theta) \equiv \mathbb{E}_{[\bm{\tilde Y} | \bm\theta]} \bm{\tilde Y}_i = \bm F \bm A$. Second, 
\begin{align*}
\mathbb{E}_{[\bm\theta | \bm Y]}\left[\hat v(\bm\theta) + \Big|\Big| \bm{\hat Y}_i(\bm\theta) -  \bm F \bm{\Delta}\bm{\tilde x}_i \Big|\Big|_2^2\right] 
&= \hat v +  \mathbb{E}_{[\bm\theta | \bm Y]}\Big|\Big| \left[\bm{\hat Y}_i(\bm\theta) - \bm{\hat Y}_i\right] +  \left[ \bm{\hat Y}_i - \bm F \bm{\Delta}\bm{\tilde x}_i \right] \Big|\Big|_2^2\\
&=  \hat v + \mathbb{E}_{[\bm\theta | \bm Y]}\Big|\Big| \bm{\hat Y}_i(\bm\theta) - \bm{\hat Y}_i \Big|\Big|_2^2 +  \Big|\Big|  \bm{\hat Y}_i - \bm F \bm{\Delta}\bm{\tilde x}_i   \Big|\Big|_2^2
\end{align*}
where $\hat v \equiv \mathbb{E}_{[\bm \theta |\bm Y]}\hat v(\bm\theta)$ and $\bm{\hat Y}_i  = \mathbb{E}_{[\bm \theta | \bm Y]} \bm{\hat Y}_i(\bm\theta) =  \mathbb{E}_{[\bm \theta | \bm Y]} \bm F \bm A\bm{\tilde x}_i = \bm F \utwi{\bar{ A}}\bm{\tilde x}_i$ for $\utwi{\bar{ A}} = \mathbb{E}[\bm{{A}} | \bm Y]$ since $\bm F$ is known. Ignoring constants that do not depend on $\bm \Delta$, we only retain the last term:
\begin{align*}
\Big|\Big|  \bm{\hat Y}_i - \bm F \bm{\Delta}\bm{\tilde x}_i   \Big|\Big|_2^2 &= \Big|\Big|  \bm F \utwi{\bar{ A}}\bm{\tilde x}_i - \bm F \bm{\Delta}\bm{\tilde x}_i   \Big|\Big|_2^2 \\
&= \Big|\Big|  \utwi{\bar{ A}}\bm{\tilde x}_i -  \bm{\Delta}\bm{\tilde x}_i   \Big|\Big|_2^2 
\end{align*}
using orthonormality of $\bm F$. The result follows immediately. 
\end{proof}

\begin{proof}[{Proof (Theorem 2)}]
Ignoring the intercept $\bm F \bm \delta_0$ as before, let $\mathcal{L}(\bm{\tilde Y}_i ,\bm \Delta | \bm F) = \big| \big| \bm{\tilde Y}_i - \bm F \bm \Delta\bm{\tilde x}_i \big| \big|_2^2$ and let $\bm \theta_{-\bm F}$ denote model parameters excluding $\bm F$. Denoting all conditional densities $p(\cdot | \cdot)$,
\begin{align*}
\mathbb{E}_{[\bm F | \bm Y]}\left\{ \mathbb{E}_{[\bm{\tilde Y} | \bm F, \bm Y]} \mathcal{L}(\bm{\tilde Y},\bm \Delta | \bm F)\right\}
&= \int p(\bm F | \bm Y) \left\{ \int p(\bm{\tilde Y} | \bm F, \bm Y) \mathcal{L}(\bm{\tilde Y},\bm \Delta | \bm F) d \bm{\tilde Y}\right\} d \bm{F}\\
&= \int p(\bm F | \bm Y) \left\{ \int \left[ \int p(\bm{\tilde Y} | \bm \theta) p(\bm \theta_{-\bm F} | \bm F, \bm Y) d \bm \theta_{-\bm F} \right] \mathcal{L}(\bm{\tilde Y},\bm \Delta | \bm F) d \bm{\tilde Y}\right\} d \bm{F}\\
&=  \int p(\bm \theta | \bm Y) \left\{ \int  p(\bm{\tilde Y} | \bm \theta)  \mathcal{L}(\bm{\tilde Y},\bm \Delta | \bm F) d \bm{\tilde Y}\right\} d \bm{\theta}\\
&= \mathbb{E}_{[\bm \theta | \bm Y]}\left\{ \mathbb{E}_{[\bm{\tilde Y} | \bm \theta]} \mathcal{L}(\bm{\tilde Y},\bm \Delta | \bm F)\right\}
\end{align*}
The remainder of the proof follows from the proof of Theorem 1, with one modification of the second step:
\begin{align*}
\mathbb{E}_{[\bm\theta | \bm Y]}\left[\hat v(\bm\theta) + \Big|\Big| \bm{\hat Y}_i(\bm\theta) -  \bm F \bm{\Delta}\bm{\tilde x}_i \Big|\Big|_2^2\right]  
&=\mathbb{E}_{[\bm\theta | \bm Y]}\left[\hat v(\bm\theta) + \Big|\Big| \bm F \bm A \bm{\tilde x}_i -  \bm F \bm{\Delta}\bm{\tilde x}_i \Big|\Big|_2^2\right]  \\
&= \mathbb{E}_{[\bm\theta | \bm Y]}\left[\hat v(\bm\theta) + \Big|\Big| \bm A \bm{\tilde x}_i -   \bm{\Delta}\bm{\tilde x}_i \Big|\Big|_2^2\right]  \\
&= \hat v +  \mathbb{E}_{[\bm\theta | \bm Y]}\Big|\Big| \left[\bm A \bm{\tilde x}_i -  \utwi{\bar{ A}} \bm{\tilde x}_i\right] +  \left[ \utwi{\bar{ A}} \bm{\tilde x}_i -  \bm{\Delta}\bm{\tilde x}_i \right] \Big|\Big|_2^2\\
&= \hat v +  \mathbb{E}_{[\bm\theta | \bm Y]}\Big|\Big|\bm A \bm{\tilde x}_i -  \utwi{\bar{ A}} \bm{\tilde x}_i\Big|\Big|_2^2 +  \Big|\Big|\utwi{\bar{ A}} \bm{\tilde x}_i -  \bm{\Delta}\bm{\tilde x}_i \Big|\Big|_2^2
\end{align*}
using orthonormality of $\bm F$ in the posterior distribution. The result follows as before. 
\end{proof}

\subsection*{Application Details}
\subsubsection*{Data Processing}


Due to substantial missingness among questionnaire responses, we model sleep questionnaire items as categorical variables. Instead of using the categorical variables from the survey directly, which would not have been as interpretable and would have left many groups with few observations, we collapsed the levels from the survey. For example, the questionnaire item \texttt{bcksleep} contributed missing, low and high factors. The low response was coded as a survey response of $1$ or $2$, high of $3$, $4$, or $5$, and missing corresponded to no response, or NA. Table \ref{table:questions} contains a description of the questionnaire items, while Table \ref{bigtable} shows the coding for all items except \texttt{wkdaysleepdur} and \texttt{bedtmwkday}, which were coded according to their corresponding level names. The full survey can be found at the National Sleep Research Resource's website under the Multi-Ethnic Study of Atherosclerosis dataset.

\begin{table}[h]
\begin{center}
\begin{tabular}{ p{1in} p{1in} p{4in} }
Item name & Factors & Question \\
\hline
  \texttt{bcksleep}      & missing/low/high                & Past 4 Weeks: Trouble Getting Back To Sleep After You Waking Too Early \\
  \texttt{bedtmwkday}    & missing/5-7/8-9/10-11/later     & Bedtime Weekday \\
  \texttt{car}           & missing/low/high                & Chance Of Dozing / Fall Asleep While: In Car Stopped In Traffic \\
  \texttt{dinner}        & low/high                        & Chance Of Dozing / Fall Asleep While: At Dinner Table \\
  \texttt{driving}       & low/high                        & Chance Of Dozing / Fall Asleep While: While Driving \\
  \texttt{extrahrs}      & missing/low/high                & Days Per Month Working Extra Hours Beyond Usual Schedule \\
  \texttt{feelngbstpk}   & missing/low/high                & Time Of Day Reaching Best Feeling Peak \\
  \texttt{feelngbstr}    & high/low                        & Time Of Day Feeling Best \\
  \texttt{insmnia}       & missing/no/yes                  & Told By Doctor As Having: Insomnia \\
  \texttt{irritable}     & low/high                        & Past 4 Weeks: Have Sleep Difficulties Causing Irritability \\
  \texttt{legsdscmfrt}   & missing/no/yes                  & Experience Desire To Move Legs Because Of Discomfort / Disagreeable Sensations In Legs \\
  \texttt{lyngdwn}       & high/low                        & Chance Of Dozing / Fall Asleep While: Lying Down To Rest In Afternoon \\
  \texttt{mosttired}     & missing/low/high                & Time In Evening Feel Most Tired And In Need Of Sleep \\
  \texttt{nap}           & missing/low/high                & Usual Week: Number Of Naps (5 Minutes Or More) \\
  \texttt{quietly}       & missing/low/high                & Chance Of Dozing / Fall Asleep While: Sitting Quietly After Lunch (No Alcohol) \\
  \texttt{readng}        & missing/low/high                & Chance Of Dozing / Fall Asleep While: Sitting And Reading \\
  \texttt{riding}        & missing/low/high                & Chance Of Dozing / Fall Asleep While: Riding As Passenger In Car \\
  \texttt{rstlesslgs}    & missing/no/yes                  & Told By Doctor As Having: Restless Legs \\
  \texttt{rubbnglgs}     & missing/no/yes                  & Feel Need To Move To Relieve Discomfort By Walking Or Rub Legs \\
  \texttt{sittng}        & missing/low/high                & Chance Of Dozing / Fall Asleep While: Sitting Inactive In Public \\
  \texttt{sleepy}        & missing/low/high                & Past 4 Weeks: Feel Overly Sleepy During Day \\
  \texttt{slpapnea}      & no/yes                          & Told By Doctor As Having: Sleep Apnea \\
  \texttt{slpngpills}    & missing/low/high                & Past 4 Weeks: Take Sleeping Pills To Help Sleep \\
  \texttt{snored}        & missing/low/high                & Past 4 Weeks: Snored \\
  \texttt{stpbrthng}     & missing/low/high                & Past 4 Weeks: Stop Breathing During Sleep \\
  \texttt{talkng}        & low/high                        & Chance Of Dozing / Fall Asleep While: Sitting And Talking To Someone \\
  \texttt{tired}         & missing/low/high                & How Tired During First Half Hour After Having Woken In Morning \\
  \texttt{trbleslpng}    & low/high                        & Past 4 Weeks: Trouble Falling Asleep \\
  \texttt{tv}            & missing/low/high                & Chance Of Dozing / Fall Asleep While: Watching TV \\
  \texttt{types}         & missing/evening/ morning/neither & Type Of Person: Morning Or Evening \\
  \texttt{typicalslp}    & missing/low/high                & Past 4 Weeks: Overall Typical Night Sleep \\
  \texttt{wakeearly}     & missing/no/yes                  & Wake up earlier than planned \\
  \texttt{wakeup}        & yes/no                          & Wake up several times a night \\
  \texttt{wkdaysleepdur} & missing/\newline\textless7/7-9/\textgreater9               & Sleep duration weekday (hours) \\
\end{tabular}
\caption{Questionnaire items included in the model and corresponding variable levels. The first level in each row is the contrasting variable.
  \label{table:questions}}
\end{center}
\end{table}

\begin{table}
  \caption{
Mapping from item levels in the model to item levels of the survey.
Excluded are \texttt{wkdaysleepdur} and \texttt{bedtmwkday} which
were coded according to the corresponding level names.
  \label{bigtable} }

\begin{minipage}{0.5\textwidth}
  \begin{tabular}{ p{0.9in} p{0.9in} p{0.9in} }
\toprule
Item name & Factors & Coding \\
\midrule
  \multirow{3}{*}{\texttt{bcksleep}}      
  & missing & NA\\
  & low     & 1,2\\
  & high    & 3,4,5\\ \hline
  \multirow{3}{*}{\texttt{car}}
  & missing & NA\\
  & low     & 1\\
  & high    & 2,3,4\\ \hline
  \multirow{2}{*}{\texttt{dinner}}
  & high    & 2,3,4\\
  & low     & 1\\ \hline
  \multirow{2}{*}{\texttt{driving}}
  & high    & 2,3,4\\
  & low     & 1\\ \hline
  \multirow{3}{*}{\texttt{extrahrs}}
  & missing & NA\\
  & low     & 0\\
  & high    & $>0$\\ \hline
  \multirow{3}{*}{\texttt{feelngbstpk}}
  & missing & NA\\
  & low     & 1,2\\
  & high    & 3,4,5\\ \hline
  \multirow{2}{*}{\texttt{feelngbstr}}
  & high    & 5\\
  & low     & 1,2,3,4\\ \hline
  \multirow{3}{*}{\texttt{insmnia}}
  & missing & NA\\
  & no      & 0\\
  & yes     & 1\\ \hline
  \multirow{2}{*}{\texttt{irritable}}
  & low     & 1\\
  & high    & 2,3,4,5\\ \hline
  \multirow{3}{*}{\texttt{legsdscmfrt}}
  & missing & NA,9 \\
  & no      & 0 \\
  & yes     & 1 \\ \hline
  \multirow{2}{*}{\texttt{lyngdwn}}
  & high & 2,3,4\\
  & low  & 1\\ \hline
  \multirow{3}{*}{\texttt{mosttired}}
  & missing & NA \\
  & low     & 1,2 \\
  & high    & 3,4,5 \\ \hline
  \multirow{3}{*}{\texttt{nap}}
  & missing & NA \\
  & low     & 0 \\
  & high    & 1 \\ \hline
  \multirow{3}{*}{\texttt{quietly}}
  & missing & NA \\
  & low     & 1 \\
  & high    & 2,3,4 \\ \hline
  \multirow{3}{*}{\texttt{readng}}
  & missing & NA \\
  & low     & 1 \\
  & high    & 2,3,4 \\ \hline
  \multirow{3}{*}{\texttt{riding}}
  & missing & NA \\
  & low     & 1 \\
  & high    & 2,3,4 \\

\bottomrule
\end{tabular}

\end{minipage} \hfill
\begin{minipage}{0.5\textwidth}
  \begin{tabular}{ p{0.9in} p{0.9in} p{0.9in} }
\toprule
Item name & Factors & Coding \\
\midrule
  \multirow{3}{*}{\texttt{rstlesslgs}}
  & missing & NA \\
  & no      & 0 \\
  & yes     & 1 \\ \hline
  \multirow{3}{*}{\texttt{rubbnglgs}}
  & missing & NA,9 \\
  & no      & 0 \\
  & yes     & 1 \\ \hline
  \multirow{3}{*}{\texttt{sittng}}
  & missing & NA \\
  & low     & 1 \\
  & high    & 2,3,4 \\ \hline
  \multirow{3}{*}{\texttt{sleepy}}
  & missing & NA \\
  & low     & 1 \\
  & high    & 2,3,4,5 \\ \hline
  \multirow{2}{*}{\texttt{slpapnea}}
  & no      & 0 \\
  & yes     & 1 \\ \hline
  \multirow{3}{*}{\texttt{slpngpills}}
  & missing & NA \\
  & low     & 1 \\
  & high    & 2,3,4,5 \\ \hline
  \multirow{3}{*}{\texttt{snored}}
  & missing & NA,9 \\
  & low     & 1 \\
  & high    & 2,3,4,5 \\ \hline
  \multirow{3}{*}{\texttt{stpbrthng}}
  & missing & NA,9 \\
  & no      & 1 \\
  & yes     & 2,3,4 \\ \hline
  \multirow{2}{*}{\texttt{talkng}}
  & no      & 1 \\
  & yes     & 2,3,4 \\ \hline
  \multirow{2}{*}{\texttt{tired}}
  & low     & 3,4,5 \\
  & high    & 1,2 \\ \hline
  \multirow{2}{*}{\texttt{trbleslpng}}
  & low     & 1 \\
  & high    & 2,3,4,5 \\ \hline
  \multirow{2}{*}{\texttt{tv}}
  & low     & 1,2 \\
  & high    & 3,4 \\ \hline
  \multirow{4}{*}{\texttt{types}}
  & missing & NA \\
  & evening & 3,4 \\
  & morning & 1,2 \\
  & neither & 5   \\ \hline
  \multirow{3}{*}{\texttt{typicalslp}}
  & missing & NA \\
  & low     & 0,1,2 \\
  & high    & 3,4,5 \\ \hline
  \multirow{3}{*}{\texttt{wakeearly}}
  & missing & NA \\
  & no  & 1,2 \\
  & yes & 3,4,5 \\ \hline
  \multirow{2}{*}{\texttt{wakeup}}
  & yes      & 3,4,5 \\
  & no     & 1,2 \\
\bottomrule
\end{tabular}
\end{minipage}
\end{table}
%
%



\subsubsection*{MCMC Diagnostics}

\begin{figure}[h]
\begin{center}
\includegraphics[width=0.85\textwidth]{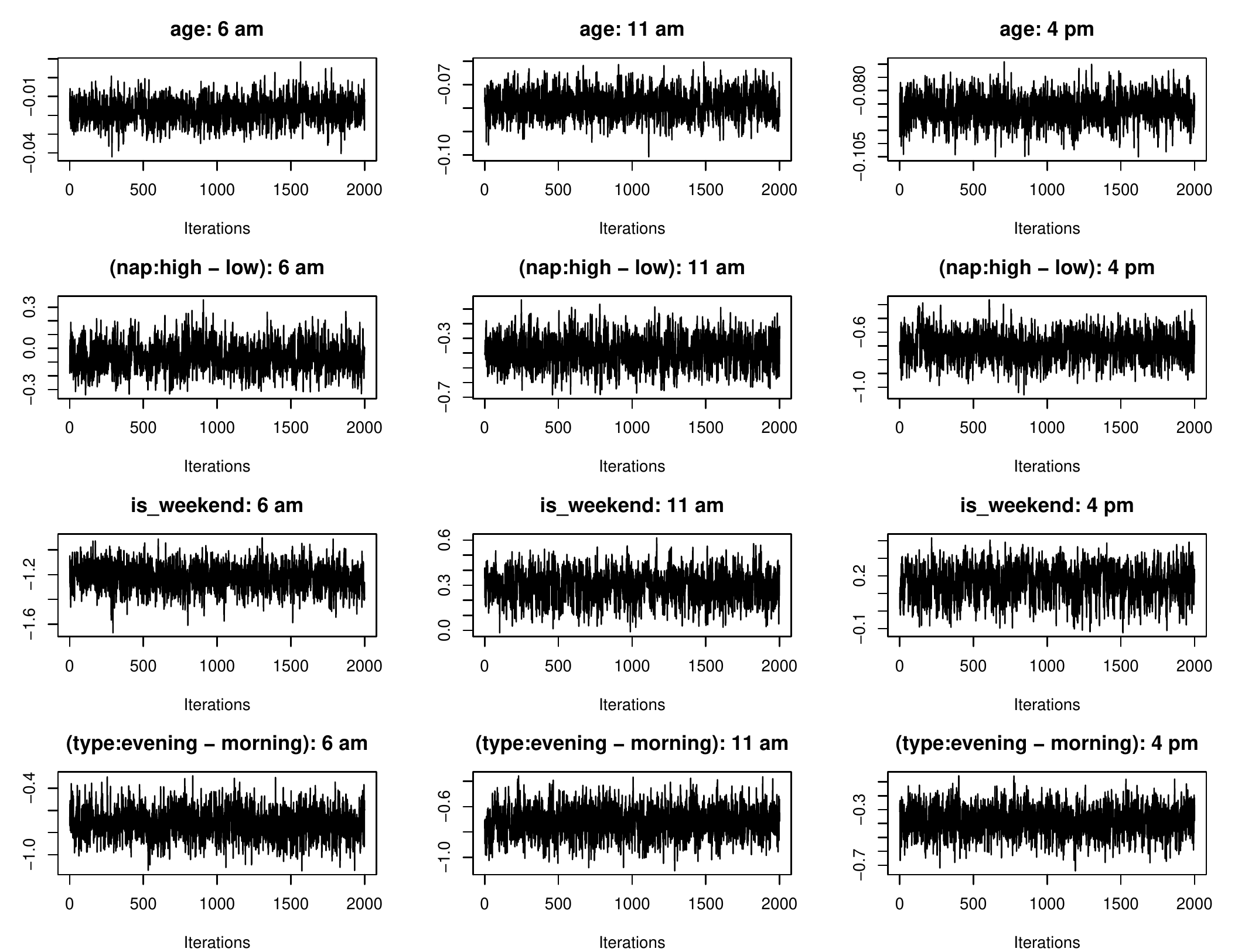}
\caption{
  Traceplot for \texttt{age}, \texttt{nap:high-low},
  \texttt{is\_weekend}, \texttt{type:evening-morning}
  regression functions evaluated at 6am, 11am and 4pm coefficient.
  \label{fig:trace}
}
\end{center}
\end{figure}

We ran the MCMC sampler for 8000 iterations. After an initial burn-in of 2000
iterations, every third sample was saved. Longer chains and dispersed starting
values did not produce noticeably different results.
The traceplots for $\tilde\alpha_j(\tau)$ with predictor $j=$ \texttt{age},
\texttt{nap:high-low}, \texttt{is\_weekend} or \texttt{type:evening-morning}
and timepoint $\tau =$ 6am, 11am or 4pm are displayed in
Figure \ref{fig:trace}.

\end{document}